\documentclass[aps,prd,preprintnumbers,nofootinbib,floatfix,11pt]{revtex4-2}

\usepackage{amsfonts,amsmath,amssymb,graphicx,epsfig,color} 

\newcommand{\be}{\begin{equation}}
\newcommand{\ee}{\end{equation}}
\newcommand{\ba}{\begin{eqnarray}}
\newcommand{\ea}{\end{eqnarray}}
\newcommand{\nn}{\nonumber}
\newcommand{\mn}{{\mu \nu}}

\newcommand{\qk}{q_{\vec{k},s}}

\begin{document}

\title{Signatures of the Quantization of Gravity at Gravitational Wave Detectors}
\author{Maulik Parikh$^{1}$, Frank Wilczek$^{2}$, and George Zahariade$^{1}$}

\affiliation{$^{1,2}$Department of Physics, Arizona State University, Tempe, Arizona 85287, USA}
\affiliation{$^{1}$Beyond Center for Fundamental Concepts in Science, Arizona State University, Tempe, Arizona 85287, USA} 
\affiliation{$^{2}$Department of Physics, Stockholm University, Stockholm SE-106 91, Sweden}
\affiliation{$^{2}$Center for Theoretical Physics, Massachusetts Institute of Technology, Cambridge, Massachusetts 02139, USA}
\affiliation{$^{2}$Wilczek Quantum Center, Department of Physics and Astronomy, Shanghai Jiao Tong University, Shanghai 200240, China}
\begin{abstract}
\begin{center}
{\bf Abstract}
\end{center}
\noindent
We develop a formalism to calculate the response of a model gravitational wave detector to a quantized gravitational field. Coupling a detector to a quantum field induces stochastic fluctuations (``noise") in the length of the detector arm. The statistical properties of this noise depend on the choice of quantum state of the gravitational field.  We characterize the noise for vacuum, coherent, thermal, and squeezed states. For coherent states, corresponding to classical gravitational configurations, we find that the effect of gravitational field quantization is small. However, the standard deviation in the arm length 
can be enhanced -- possibly significantly -- when the gravitational field is in a non-coherent state. The detection of this fundamental noise could provide direct evidence for the quantization of gravity and for the existence of gravitons. 

\end{abstract}

\maketitle

\section{Introduction}

\noindent

The present relationship between general relativity and quantum mechanics is ironic. On the one hand, a fully quantum-mechanical treatment of gravity raises deep conceptual issues, which come to a head in the treatment of black hole evaporation and early-universe cosmology.  On the other hand, general relativity itself can be derived from consistency conditions on the quantum theory of a massless helicity-two particle: the  graviton~\cite{Feynman:1963ax,Weinberg:1964kqu,Deser:1969wk,Boulware:1974sr}.  Finally, all existing experiments and observations in physics, including many in which both gravity and quantum mechanics play central roles, have been described successfully within a semi-classical theory, wherein the gravitational field can be treated classically; experimentally, we have hardly any evidence at all that gravity is quantized. (The detection of B-mode polarization in the cosmic microwave background though would have provided indirect evidence for the quantization of gravity \cite{Krauss:2014sua}.)   

With the discovery of gravitational waves, it is of paramount interest to examine possible implications of the quantization of gravity for gravitational wave detectors, such as LIGO~\cite{Abbott:2016blz} or LISA~\cite{Audley:2017drz}. Several authors have proposed that classical treatment of the gravitational field might not be wholly adequate in this context~\cite{AmelinoCamelia:1998ax,AmelinoCamelia:1999gg,Verlinde:2019xfb,Verlinde:2019ade,Lieu:2017lzh,Guerreiro:2019vbq}, based on possible inadequacies of general relativity or on intuition about graviton shot noise. In contrast, Dyson~\cite{Dyson:2013jra} has argued that since one has only barely detected gravitational waves, and since a typical gravitational wave has of order $10^{37}$ gravitons within a cubic wavelength, one would have to increase detector sensitivity by some 37 orders of magnitude in order to discern the discrete character of gravitons. Extending Dyson's conclusion, table-top approaches to detecting gravitons directly have also been regarded as unpromising~\cite{Rothman:2006fp,Boughn:2006st}. These arguments, which have been largely heuristic, have thus led to inconsistent predictions about the possible observable signatures of quantum gravity.

Here we present a formalism for rigorously computing the effects of the quantization of the gravitational field on gravitational wave interferometers. We will treat the gravitational field as a quantum-mechanical entity, and bring in its quantum mechanics perturbatively. This allows us to get definite equations and assess the quantitative importance of quantum gravity effects whose existence seems theoretically secure. Our main finding is that coupling to a quantized gravitational field induces fluctuations, or noise, in the length of the arm of a gravitational wave interferometer. The noise, which appears to be correlated between nearby detectors, has statistical properties that depend on the quantum state of the gravitational field. The quantum state in turn depends on the sources of gravity. Within this framework we {\em derive} the result that for a wide range of gravitational sources the deviations from classical behavior are expected to be minuscule, but we also identify some plausible exceptions. This paper supplements and extends two shorter works~\cite{Parikh:2020nrd,Parikh:shortpaper}. 

An outline of this paper is as follows. We begin, in Section~\ref{classaction}, by introducing a simple model of a gravitational wave detector, or ``arm" for short. Our model detector consists of two free-falling masses whose geodesic separation is being monitored. Decomposing the gravitational field into modes leads to an action for each mode, \eqref{Somega}, which describes a simple harmonic oscillator coupled to a free particle via a Yukawa-type (cubic) derivative interaction. In Section~\ref{qmdet}, we consider the quantum mechanics of this system. More specifically, we employ the Feynman-Vernon influence functional method~\cite{Feynman:1963fq}, which enables one to determine the effect, or influence, of one quantum subsystem on another. (An alternative approach is considered in~\cite{Kanno:2020usf}.) This technique has been used extensively in the literature to study dissipation in open systems, the semi-classical limit of quantum field theories, as well as within the field of stochastic gravity~\cite{Hu:1999mm,Hu:1991di,Hu:1993vs,Calzetta:1993qe,Johnson:2000if,Johnson:2000qd}. In our context, it yields the effect of a single gravitational mode on the physics of the detector arm length. The result of this quantum-mechanical calculation is the influence functional, (\ref{geninffunc}). We find that the influence functional generically factorizes into a ground state component and a piece that depends on the quantum state of the mode. In Section~\ref{qmfield}, we extend our calculation to quantum field theory by summing over all gravitational modes; the sum depends on the choice of quantum state of the gravitational field. Several different states are considered: the vacuum state, a coherent state corresponding to a quantized gravitational wave, a thermal density matrix due to a cosmic background or an evaporating black hole, and a squeezed state potentially originating in certain inflationary scenarios. For each gravitational field state, we perform the mode sum with the goal of obtaining the field-theoretic influence functional. In Section~\ref{langevin}, we derive our main result: an effective equation of motion for the length of the detector arm, \eqref{langevineq}. This turns out to be a Langevin-like stochastic differential equation, as one would naturally expect: coupling a classical system to a quantum system forces its dynamics to be governed by a stochastic -- rather than a deterministic -- equation. Our Langevin equation contains three different types of source terms. First, there is a coupling of the arm to any extant classical gravitational wave. Second, there is a fifth-derivative term that corresponds to the gravitational counterpart of the Abraham-Lorentz radiation reaction force. Both of these are essentially classical. But it is the third term that is the most interesting. We find that there are fluctuations in the length of the detector arm which are due to quantum noise: noise that originates in the underlying quantum nature of the gravitational field. The statistical characteristics of the noise depend on the quantum state of the field. In Section~\ref{discussion}, we estimate the amplitude of the jitters in the arm length for various states. For coherent states (which are the quantum counterparts of classical field configurations, such as gravitational waves), we find indeed that, although the fluctuations are many orders of magnitude larger than Dyson's rough estimate, they are still unmeasurably small. But the fluctuations can be enhanced for other states of the gravitational field. In particular, for squeezed states, the enhancement can be exponentially large in the squeezing parameter, with the precise magnitude of the enhancement dependent on details of the squeezing. We conclude, in Section~\ref{summary}, with a brief summary.

\section{The Classical Action}
\label{classaction}

\noindent
Let us begin by obtaining a classical action for a weak gravitational field coupled to a model gravitational wave detector. We will explicitly retain $\hbar$ and $G$ in our expressions; the speed of light is set to one. Our metric convention is to use mostly plus signature. Consider then a weak gravitational field. We can find coordinates for which the metric can be written as
\be
g_{\mn}=\eta_{\mn} + h_{\mn} \; ,
\ee
where $\eta_{\mn} = {\rm diag}(-1,1,1,1)$ is the usual Minkowski metric in Cartesian coordinates. To quadratic order in $h_{\mu\nu}$, the Einstein-Hilbert action is
\be
S_{\rm EH}= \frac{1}{64 \pi G}\int d^4x\,\left(h_{\mn}\square h^{\mn}+2h^{\mn}\partial_\mu\partial_\nu h-h\square h-2h_{\mn}\partial_\rho\partial^\mu h^{\nu\rho}\right) \; .
\ee
Here the linear part of the action in $h_{\mn}$ has been discarded because it is a total derivative. This action inherits two sets of symmetries from the diffeomorphism invariance of Einstein's theory: (i) global Poincar\'e invariance, $x^\mu \to \Lambda^\mu_{\, \, \nu} x^\nu + a^\mu$ of the background, and (ii) gauge symmetry of the perturbation, $h_{\mn} \to h_{\mn} + \partial_\mu \xi_\nu + \partial_\nu \xi_\mu$. 
Going to the transverse-traceless (TT) gauge, the metric perturbation obeys
\ba
&\partial^\mu \bar{h}_{\mu\nu}=0\ ,\\
&u^\mu \bar{h}_{\mu\nu}=0\ ,\\
&\bar{h}^\mu{}_\mu=0 \; ,
\ea
where the bar on $h_{\mu\nu}$ signifies that we are in TT gauge. Here $u^\mu$ is an arbitrary constant time-like vector; we use a background Lorentz transformation to align the time direction so that $u^\mu=\delta^\mu_{\, \, 0}$. With these choices, the action in TT gauge reads
\be
S_{\rm EH}=-\frac{1}{64 \pi G}\int d^4x\ \partial_\mu \bar{h}_{ij}\partial^\mu \bar{h}^{ij} \; , \label{EH}
\ee
where Latin indices denote spatial directions.

Next we would like to include an action for a gravitational wave detector. It is easiest to imagine this as a pair of free-falling massive test particles, as might be the case for a pair of satellites in orbit. The geodesic separation between the two particles is then a gauge-invariant quantity, and we have in mind that there is some way of measuring that separation. Let the (comoving) TT-gauge coordinates of the two particles be $X^\mu(t)$ and $Y^\mu(t)$. Then their action is
\be
S_{\rm detector} = - M_0 \int dt \sqrt{-g_\mn (X) \dot{X}^\mu \dot{X}^\nu} - m_0 \int dt \sqrt{-g_\mn (Y) \dot{Y}^\mu \dot{Y}^\nu} \; , \label{detectoractions}
\ee
where dotted quantities are differentiated with respect to coordinate time, $t$. We have taken the particles to have different test masses $M_0$, $m_0$; since we are interested in their relative motion, we assume for convenience that $M_0\gg m_0$ and that the first particle is on-shell with worldline $X_0^\mu (t)$. Furthermore, and without loss of generality, we can place the first particle at rest at the origin of our coordinate system, $X_0^\mu(t) = t \delta^\mu_0$, so that the coordinate time $t$ is the proper time of the first particle; since $\bar{h}_{0 \mu} = 0$ in our gauge this worldline is indeed a geodesic. In this parametrization $Y^0(t)=t$. We can then make a change of variables from $Y^i$ to $\xi^i$ as follows:
\be
Y^i - X_0^i = \xi^i - \frac{1}{2} \delta^{ij} \bar{h}_{jk} (Y) \xi^k \; . \label{changetoxi}
\ee
We also assume that the separation of the two particles is less than the characteristic scale of variation of $\bar{h}_{ij}$; this is analogous to the dipole approximation in electrodynamics. In our context, this will mean that we will consider only those wavelengths that are greater than the separation of the masses. Then $\bar{h}_{ij}(Y) \approx \bar{h}_{ij}(X_0)$. We then take the non-relativistic limit so that the action becomes
\be
S_{\rm detector} = \int dt \frac{1}{2} m_0 (\delta_{ij} + \bar{h}_{ij}(X_0)) \dot{Y}^i \dot{Y}^j \; ,
\label{refdipole}
\ee
where we have dropped all non-dynamical terms. Inserting (\ref{changetoxi}), we find to lowest (linear) order in $\bar{h}_{ij}$, that
\be
S_{\rm detector} = \int dt \frac{1}{2} m_0 \left (\delta_{ij} \dot{\xi}^i \dot{\xi}^j -\dot{\bar{h}}_{ij} \dot{\xi}^i \xi^j \right ) \; . \label{xiaction}
\ee
Via an integration by parts, the second term in the Lagrangian can be written more symmetrically as $+\frac{1}{4} m_0 \ddot{\bar{h}}_{ij} \xi^i \xi^j$. 

We can think of $(t, \xi^i)$ as the coordinates of the second particle in an orthonormal non-rotating Cartesian coordinate system whose spatial origin moves with the first particle. Indeed, these are simply Fermi normal coordinates defined with respect to the worldline of the first particle. With this observation, we can easily re-derive the detector action. Denoting Fermi normal coordinate indices with hats, we can write the metric as
\ba
g_{\hat{0}\hat{0}}(t,\xi)&=&-1-R_{\hat{i}\hat{0}\hat{j}\hat{0}} (t,0) \xi^{\hat{i}}\xi^{\hat{j}} + O(\xi^3) \nn \ ,\\
g_{\hat{0}\hat{i}}(t,\xi)&=&-\frac{2}{3}R_{\hat{0}\hat{j}\hat{i}\hat{k}} (t,0) \xi^{\hat{j}}\xi^{\hat{k}}+O(\xi^3) \nn \ ,\\
g_{\hat{i}\hat{j}}(t,\xi)&=&\delta_{\hat{i}\hat{j}}-\frac{1}{3}R_{\hat{i}\hat{k}\hat{j}\hat{l}} (t,0) \xi^{\hat{k}}\xi^{\hat{l}}+O(\xi^3)\ ,
\ea
where the Riemann tensor has been evaluated at $X_0^\mu(t) = (t,0)$. We can now use the fact that, to first order in the metric perturbation, $R_{\hat{i}\hat{0}\hat{j}\hat{0}}(t,0)=R_{{i}{0}{j}{0}}(t,0)$, where the unhatted indices correspond to TT gauge~\cite{Misner:1974qy}. Then
\be
R_{\hat{i}\hat{0}\hat{j}\hat{0}}(t,0) = - \frac{1}{2} \ddot{\bar{h}}_{ij}(t,0)\ .
\ee
Picking $Y^\mu=(t,\xi^{\hat{i}})$ and inserting into (\ref{detectoractions}), we recover (\ref{xiaction}) in the appropriate limit. We see that, technically, the indices on $\xi$ in~\eqref{changetoxi} and~\eqref{xiaction} should be hatted; the change of variables~\eqref{changetoxi} can be interpreted as a switch from the coordinate separation $Y^i$ to the physical separation $\xi^{\hat{i}}$. 

Next, we decompose $\bar{h}_{ij}$ into discrete modes:
\be
\bar{h}_{ij}(t,\vec{x}) = \frac{1}{\sqrt{\hbar G}} \sum_{\vec{k},s} \qk(t) e^{i \vec{k} \cdot \vec{x}} \epsilon^s_{ij} (\vec{k})\ . 
\label{hdecomp}
\ee
Here $\qk$ is the mode amplitude. The discreteness of the decomposition (\ref{hdecomp}) can be achieved, for example, by working in a cubic box of side $L$, so that the wave vectors are $\vec{k} = 2 \pi \vec{n}/L$ with $\vec{n} \in \mathbb{Z}^3$. The label $s = +, \times$ indicates the polarization, and $\epsilon^s_{ij}$ is the polarization tensor, satisfying normalization, transversality, and tracelessness conditions:
\ba
\epsilon^s_{ij}(\vec{k}) \epsilon^{ij}_{s'}(\vec{k}) & = & 2 \delta^s{}_{s'}\ , \\
k^i \epsilon^s_{ij} (\vec{k}) & = & 0\ , \\
\delta^{ij} \epsilon^s_{ij} (\vec{k}) & = & 0\ .
\ea
In finite volume, the orthonormality of the Fourier modes means
\be
\int d^3x \, e^{i (\vec{k} - \vec{k}') \cdot \vec{x}} = L^3 \delta_{\vec{k}, \vec{k}'}\ ,
\ee
where $\delta_{\vec{k}, \vec{k}'}$ is a Kronecker delta.
Inserting (\ref{hdecomp}) into (\ref{EH}) and (\ref{xiaction}), we find
\be
S = \frac{L^{3}}{32 \pi\hbar G^2} \int dt \sum_{\vec{k},s} \left ( | \dot{q}_{\vec{k},s} |^{ 2} - \vec{k}^2  | \qk  |^{2} \right ) +\int dt \frac{1}{2} m_0 \left (\delta_{ij} \dot{\xi}^i \dot{\xi}^j -  \frac{1}{\sqrt{\hbar G}} \sum_{\vec{k},s} \dot{q}_{\vec{k},s} \epsilon^s_{ij}(\vec{k}) \dot{\xi}^i \xi^j \right ) \ .
\ee
The reality of $\bar{h}_{ij}$ implies that
\be
\qk^* \epsilon^s_{ij}(\vec{k})  = q_{-\vec{k},s} \epsilon^s_{ij}(-\vec{k})\ . 
\ee
Using this reality condition, we have
\be
\sum_{\vec{k},s} \dot{q}_{\vec{k},s} \epsilon^s_{ij}(\vec{k}) \dot{\xi}^i \xi^j = \frac{1}{2}\sum_{\vec{k},s} (\dot{q}_{\vec{k},s} + \dot{q}_{\vec{k},s}^*)\epsilon^s_{ij}(\vec{k}) \dot{\xi}^i \xi^j = \sum_{\vec{k},s} \left ( {\rm Re} \, \dot{q}_{\vec{k},s} \right ) \epsilon^s_{ij}(\vec{k}) \dot{\xi}^i \xi^j\ .
\ee
Evidently, only the real part of the mode amplitude couples to the detector; we therefore discard the imaginary part and take $\qk$ hereafter to be real. Defining
\be
m \equiv \frac{L^{3}}{16 \pi \hbar G^2}\,,
\label{mdef}
\ee
we obtain
\be
S =  \int dt \sum_{\vec{k},s} \frac{1}{2} m \left (  \dot{q}_{\vec{k},s}^{ 2} - \vec{k}^2  \qk^2 \right ) +\int dt \frac{1}{2} m_0 \left (\delta_{ij} \dot{\xi}^i \dot{\xi}^j - \frac{1}{\sqrt{\hbar G}} \sum_{\vec{k},s} \dot{q}_{\vec{k},s} \epsilon^s_{ij}(\vec{k}) \dot{\xi}^i \xi^j \right )\ .  
\label{genaction}
\ee
Now consider a single mode with wave vector $\vec{k}$ directed along the positive $z$-axis and with magnitude $\omega=|\vec{k}|$. Restricting to the $+$ polarization for simplicity, and dropping the subscripts on $q_{\vec{k},s}$, the action for this mode reduces to
\be
S_\omega =  \int dt \left( \frac{1}{2} m (\dot{q}^{ 2} - \omega^2  q^2) + \frac{1}{2} m_0 \left(\dot{\xi}_x^2 + \dot{\xi}_y^2+ \dot{\xi}_z^2 - \frac{1}{\sqrt{\hbar G}} \dot{q} (\dot{\xi}_x\xi_x - \dot{\xi}_y\xi_y)\right) \right) \,. 
\ee
Let us orient the $x$-axis to coincide with the line joining the two test masses at time $t = 0$ so that $\xi_y(0)=\xi_z(0)=0$. Since the masses are initially at rest with respect to each other, we have $\dot{\xi}_x(0)=\dot{\xi}_y(0)=\dot{\xi}_z(0)=0$. With this initial condition, we see that $\xi_y$ and $\xi_z$ are not excited by the gravitational wave mode at all and hence $\xi_y(t) = \xi_z (t) = 0$ on shell. (Quantum mechanically, $\xi_y$ and $\xi_z$ could still fluctuate but we ignore this for simplicity.) Dropping the subscript on $\xi_x$, 
and defining
\be
g \equiv \frac{m_0}{2\sqrt{\hbar G}}\,,
\label{gdef}
\ee
we finally arrive at
\be
S_\omega = \int dt \left ( \frac{1}{2} m (\dot{q}^{ 2} - \omega^2  q^2) + \frac{1}{2} m_0  \dot{\xi}^2 - g \dot{q} \dot{\xi} \xi  \right ) \, .\label{Somega}
\ee 
We have found an action for a gravitational mode of energy $\hbar\omega$, with amplitude proportional to $q$, interacting with a free-falling mass $m_0$ whose geodesic separation (``arm length'') from a heavier fixed mass is given by $\xi$.
This action corresponds to a simple harmonic oscillator coupled to a free particle via a cubic derivative interaction. Let us quantize it.

\section{Quantum Mechanics of the Mode-Detector System}
\label{qmdet}

\noindent
Our aim is to investigate the effect of the quantization of the gravitational field on the arm length $\xi$ of a model gravitational wave detector. Given a specified initial state of the gravitational field, and summing over its unknown final states, the most general quantity one can calculate is the transition probability between two states of $\xi$, $\phi_A$ and $\phi_B$, within a finite time interval $T$. We hasten to add, however, that we will ultimately regard the detector arm as classical, and we will use our formula for the transition probability mainly to extract the quantum-corrected equation of motion for $\xi$. Determining the transition probability calls for a quantum field theory calculation with the action given by the continuum limit of~\eqref{genaction}. In this section, as a stepping stone, we shall consider the quantum mechanics of just a single mode. Later, in Section~\ref{qmfield}, we will extend our results to field theory by summing over a continuum of modes.

The calculation of transition probabilities for $\xi$ in the presence of a single mode of the gravitational field in some specified initial state is a problem in ordinary quantum mechanics. It can be solved analytically. Nonetheless, the derivation is lengthy and brings in several subtleties, and involves aspects of quantum mechanics that may be unfamiliar to many physicists. 

The primary object of interest is the Feynman-Vernon influence functional~\cite{Feynman:1963fq}, which is a powerful tool for determining the complete dynamics of a  quantum system interacting with another unobserved quantum system. The final expressions for this are~\eqref{geninffunc}, \eqref{W}, \eqref{vacinffunc}. 

\begin{figure}[ht]
\centering
\includegraphics[width=0.7\textwidth]{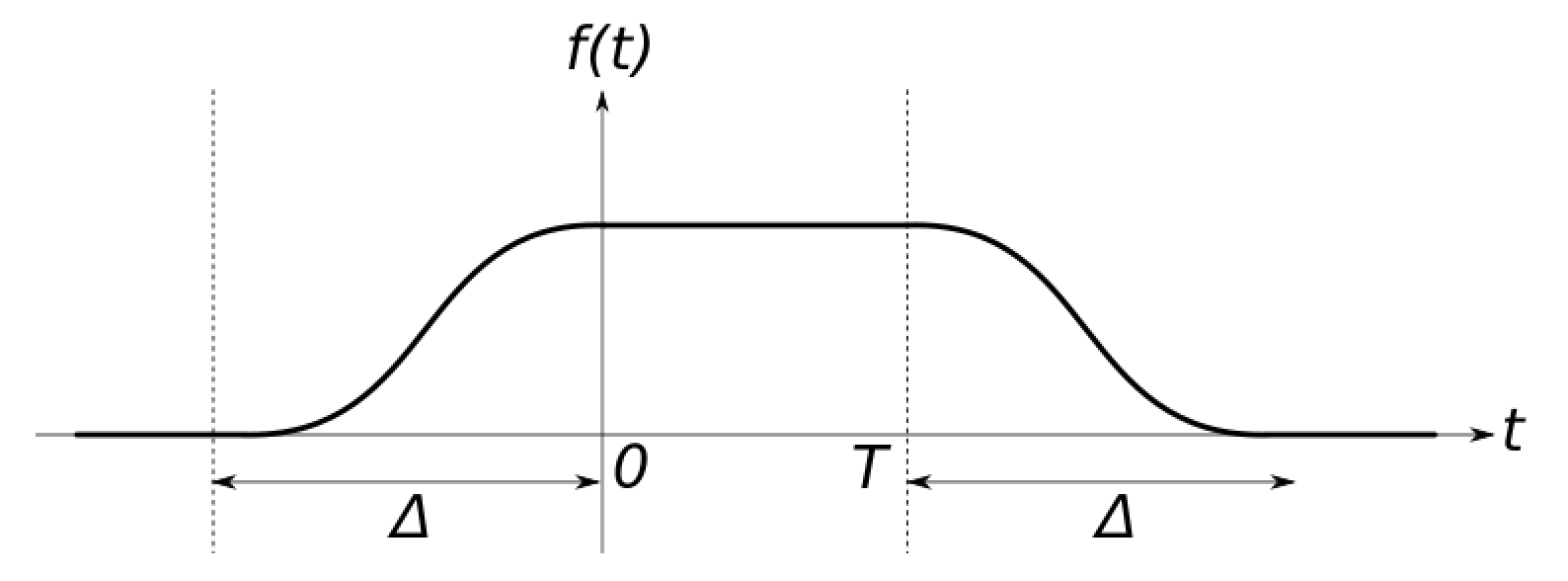}
\caption{Switching on and off function}
\label{switch}
\end{figure}

The classical dynamics of the single mode is given by~\eqref{Somega}, which describes a quantum harmonic oscillator, $q(t)$, coupled to a free particle, $\xi(t)$. (Recall that $\xi(t)$ is the length of the detector arm.) We will quantize both $q$ and $\xi$ but we expect that $\xi$ will ultimately be well-approximated as classical. Let us introduce the canonical momenta
\ba
p&=&m\dot{q}-g\dot{\xi}\xi\,,\\
\pi&=&m_0\dot{\xi}-g\dot{q}\xi\,,
\ea
conjugate to the variables $q$ and $\xi$, respectively. The Hamiltonian then reads
\be
H(q,p,\xi,\pi)=\left(\frac{p^2}{2m}+\frac{\pi^2}{2m_0}+\frac{gp\pi\xi}{mm_0}\right)\left(1-\frac{g^2\xi^2}{mm_0}\right)^{\! -1}+\frac{1}{2}m\omega^2q^2\,.\label{fullHam}
\ee
This Hamiltonian contains a cubic interaction term coupling two momenta and a position, as well as an overall non-polynomial position-dependent factor multiplying the momentum-dependent terms. Nevertheless, as we will see, we will be able to obtain some exact expressions. Notice that for $g = 0$, the Hamiltonian reduces to that of two decoupled degrees of freedom:
\be
H \rightarrow \frac{p^2}{2m}+\frac{1}{2}m\omega^2q^2 + \frac{\pi^2}{2m_0} \,.
\ee
To quantize~\eqref{fullHam} we promote the positions and momenta to operators. There is formally an ordering ambiguity which we circumvent by assuming Weyl-ordering. We will also assume that the coupling $g$ is adiabatically switched on and off, $g \rightarrow f(t) g$, where $f(t)$ is a function satisfying $f(t \leq -\Delta) =f(t\geq T+\Delta)= 0$ and $f(T\geq t \geq 0) = 1$, and $\Delta$ is some time-scale that will play no role (see Fig.~\ref{switch}).

We assume that, at $t = -\infty$, the combined state of the harmonic oscillator and particle system is a tensor product state. The justification for this is that the gravitational field is created before the interaction is switched on and therefore the detector and mode are initially uncoupled. Then, in Schr\"odinger picture, the Hamiltonian evolves the harmonic oscillator and particle states independently until the interaction is switched on at time $t = - \Delta$. This means that, at time $t = - \Delta$, the combined state is still a tensor product state. We switch to Heisenberg picture at time $t = - \Delta$, when we define the harmonic oscillator state to be $|\psi_\omega \rangle$ and the particle state to be $|\phi_A \rangle$. The subscript $\omega$ on the harmonic oscillator state reminds us that it is the state of the graviton mode of energy $\hbar \omega$. Technically, other quantities should also have an $\omega$ subscript to indicate that they pertain to this particular mode, but we will omit such subscripts to reduce clutter.

We are interested in calculating the transition probability for the particle to be found in a state $|\phi_B\rangle$ at time $t = T+\Delta$ with an interaction that takes place between $t = 0$ and $t = T$. We are not interested in the final state $|f\rangle$ of the harmonic oscillator, which generically will be different from its initial state $|\psi_\omega\rangle$. Indeed, in terms of the original gravitational problem, the detector masses will typically both absorb and emit gravitons (through spontaneous as well as stimulated emission). Thus the goal of this section is to calculate 
\be
P_{\psi_\omega}(\phi_A \to \phi_B)  =  \sum_{|f\rangle} |\langle f, \phi_B | \hat{U}(T+\Delta,-\Delta) | \psi_\omega, \phi_A \rangle |^2\,,
\ee
for a given initial state, $|\psi_\omega \rangle$, of the harmonic oscillator. Here, our notation for tensor product states of the joint Hilbert space is
\be
|a,b\rangle \equiv |a\rangle\otimes|b\rangle\,,
\ee
and $\hat{U}$ is the unitary time-evolution operator associated with the Hamiltonian~\eqref{fullHam}. 

We now insert several complete bases of joint position eigenstates, $\int d q d \xi  |q, \xi \rangle \langle q, \xi|$. Then
\ba
& & P_{\psi_\omega}(\phi_A \to \phi_B) \nn \\
& = & \sum_{|f\rangle} \langle \psi_\omega, \phi_A  | \hat{U}^\dagger(T+\Delta,-\Delta) | f , \phi_B \rangle \langle f, \phi_B | \hat{U}(T+\Delta,-\Delta) | \psi_\omega, \phi_A \rangle \nn \\
& = & \sum_{|f\rangle} \int dq_i dq'_i dq_f dq'_f d \xi_i d \xi'_i d \xi_f d \xi'_f \,\langle \psi_\omega, \phi_A | q'_i, \xi'_i \rangle \langle q'_i , \xi'_i | \hat{U}^\dagger(T+\Delta,-\Delta) | q'_f , \xi'_f \rangle \langle q'_f, \xi'_f | f , \phi_B \rangle \times \nn \\
& & \hspace{+0.6cm} \langle f , \phi_B | q_f , \xi_f \rangle \langle q_f , \xi_f | \hat{U}(T+\Delta,-\Delta) | q_i , \xi_i \rangle \langle q_i, \xi_i | \psi_\omega, \phi_A \rangle \nn \\
& = & \int dq_i dq'_i dq_f d \xi_i d \xi'_i d \xi_f d \xi'_f \, \psi_\omega^*(q'_i) \phi^*_A(\xi'_i)  \phi_B(\xi'_f) \phi^*_B(\xi_f) \psi_\omega(q_i) \phi_A(\xi_i) \times
 \nn \\
& & \hspace{+0.6cm} 
\langle q'_i , \xi'_i | \hat{U}^\dagger(T+\Delta,-\Delta) | q_f , \xi'_f \rangle \,
 \langle q_f , \xi_f | \hat{U}(T+\Delta,-\Delta) | q_i , \xi_i \rangle \,.
 \label{baseprob}
\ea
Here $\psi_\omega(x)$, $\phi_A(x)$, $\phi_B(x)$ are the wave functions for the harmonic oscillator and the free particle in position representation in the states $|\psi_\omega \rangle$, $|\phi_A \rangle$, $|\phi_B\rangle$, respectively.
Next we can express each of the amplitudes in canonical path-integral form:
\be
\langle q_f , \xi_f | \hat{U}(T+\Delta,-\Delta) | q_i , \xi_i \rangle = \int {\cal D} \pi {\cal D} \xi {\cal D} p {\cal D} q \exp \left (\frac{i}{\hbar} \int_{-\Delta}^{T+\Delta} dt \left ( \pi \dot{\xi} + p \dot{q} - H(q,p,\xi,\pi) \right ) \right )\,.
\ee
Performing the path integral over $\pi$ (which has the same effect as the partial Legendre transform used to obtain  the Routhian), we find
\be
\langle q_f , \xi_f | \hat{U}(T+\Delta,-\Delta) | q_i , \xi_i \rangle = \int \tilde{\cal D} \xi e^{\frac{i}{\hbar} \int dt \frac{1}{2} m_0 \dot{\xi}^2} \int {\cal D} p {\cal D} q \exp \left (\frac{i}{\hbar} \int_{-\Delta}^{T+\Delta} dt \left (p \dot{q} - H_\xi(q,p) \right ) \right )\ ,
\label{amppi}
\ee
where
\be
H_\xi(q,p) \equiv \frac{(p + g \xi \dot{\xi})^2}{2m} + \frac{1}{2} m \omega^2 q^2 \ . \label{Hxi}
\ee
In~\eqref{amppi}, $\tilde{\cal D} \xi$ is a measure in which a $g$- and $\xi$-dependent piece has been absorbed; since ultimately we will only be interested in a saddle point of the $\xi$ path integral, we can safely disregard the details of this modified measure.

Now, the path integrals over $p$ and $q$ can themselves be thought of as giving an amplitude for the harmonic oscillator coupled to an external field, $\xi(t)$, and evolved via the Hamiltonian (\ref{Hxi}). Thus
\be
\int {\cal D} p {\cal D} q \exp \left (\frac{i}{\hbar} \int_{-\Delta}^{T+\Delta} dt \left (p \dot{q} - H_\xi(q,p) \right ) \right ) = \langle q_f | \hat{U}_\xi(T+\Delta,-\Delta) | q_i \rangle\ ,
\ee
where $\hat{U}_\xi$ is the unitary time-evolution operator associated with the Hamiltonian~\eqref{Hxi}. Then, after integration over $q_f$ in~\eqref{baseprob}, we find
\be
P_{\psi_\omega}(\phi_A \to \phi_B)
\equiv \int d \xi_i d \xi'_i d \xi_f d \xi'_f \, \phi^*_A(\xi'_i)  \phi_B(\xi'_f) \phi^*_B(\xi_f) \phi_A(\xi_i) \hspace{-15mm}\underset{\substack{\xi(-\Delta) = \xi_i \; , \; \xi'(-\Delta) = \xi'_i \\ \xi(T+\Delta) = \xi_f \; , \; \xi'(T+\Delta) = \xi'_f}} \int \hspace{-15mm} \tilde{\cal D}\xi \tilde{\cal D}\xi' e^{\frac{i}{\hbar} \int_{-\Delta}^{T+\Delta} dt \frac{1}{2} m_0 (\dot{\xi}^2 - \dot{\xi}^{'2})} F_{\psi_\omega}[\xi,\xi']\ , 
\label{nextprob}
\ee
where
\be
F_{\psi_\omega}[\xi, \xi'] = \langle \psi_\omega | \hat{U}_{\xi'}^\dagger (T+\Delta, -\Delta) \hat{U}_\xi (T+\Delta,-\Delta) | \psi_\omega \rangle\ ,
\label{ifbasic}
\ee
is the Feynman-Vernon influence functional~\cite{Feynman:1963fq}.  The influence functional encodes the entirety of the effect of coupling to the harmonic oscillator $q$ on the particle $\xi$; indeed, in~\eqref{nextprob}, the only dependence on the harmonic oscillator state $|\psi_\omega\rangle$ occurs through the influence functional. In our context, the influence functional tells us about the effect of the quantized gravitational field mode on the arm length of the detector. Significantly, as we shall see later, the coupling to quantum degrees of freedom induces stochastic fluctuations in the length of the arm, whose statistical properties can be extracted from the influence functional.

It will often be useful to work directly with the influence phase, $\Phi_{\psi_\omega}[\xi,\xi']$, defined by
\be
F_{\psi_\omega}[\xi,\xi']\equiv e^{i\Phi_{\psi_\omega}[\xi,\xi']}\,.
\ee
To gain some appreciation of the influence phase, suppose $\Phi_{\psi_\omega}[\xi, \xi']$ were to decompose additively into parts that depended separately on $\xi$ and $\xi'$, say $\Phi_{\psi_\omega}[\xi,\xi'] = (S_{\psi_\omega}[\xi] - S_{\psi_\omega}[\xi'])/\hbar$. Then, from (\ref{nextprob}), we see that the sole effect of the quantized gravitational field mode would be to add a piece $S_{\psi_\omega}[\xi]$ to the action for $\xi$. Moreover, the path integrals for $\xi$ and $\xi'$ would then decouple. However, as we shall see, the influence phase does not decompose in this way in general.

\subsection*{Evaluating the Influence Functional}

\noindent
Now we would like to obtain a more explicit expression for the influence functional~\eqref{ifbasic}. To do so, we split the time-evolution operator, $\hat{U}_\xi (T+\Delta,-\Delta) = \hat{U}_\xi (T+\Delta,T)\hat{U}_\xi (T,0) \hat{U}_\xi (0,-\Delta)$. During the switching on and off of the interaction, we invoke the adiabatic theorem to compute the effect of $\hat{U}_\xi (0,-\Delta)$ and $\hat{U}_\xi (T+\Delta,T)$ on state vectors; this means that, as the interaction is switched on, eigenstates of the Hamiltonian remain instantaneous eigenstates. But notice from the form of~\eqref{Hxi} that the instantaneous eigenstates are merely those of a simple harmonic oscillator shifted in momentum space: $p \to p + g \xi \dot{\xi}$. Since shifts in momentum space are generated by the position operator, we infer that
\ba
\hat{U}_\xi(0, -\Delta) &=& e^{-\frac{i}{\hbar} \hat{q} g \xi(0) \dot{\xi}(0)} e^{-\frac{i}{\hbar} \hat{H}_0 \Delta}\,,\\
\hat{U}_\xi(T+\Delta, T) &=& e^{-\frac{i}{\hbar} \hat{H}_0 \Delta}e^{+\frac{i}{\hbar} \hat{q} g \xi(T) \dot{\xi}(T)} \,.
\ea
Note that there is no geometric phase here. Futher, for the sake of clarity, we redefine our Heisenberg state via $e^{-\frac{i}{\hbar} \hat{H}_0 \Delta} |\psi_\omega \rangle \rightarrow  |\psi_\omega \rangle$. We therefore have
\be
F_{\psi_\omega}[\xi, \xi'] = \langle \psi_\omega | e^{\frac{i}{\hbar} \hat{q} g \xi'(0) \dot{\xi}'(0)} \hat{U}_{\xi'}^\dagger (T, 0)e^{-\frac{i}{\hbar} \hat{q} g \xi'(T) \dot{\xi}'(T)} e^{\frac{i}{\hbar} \hat{q} g \xi(T) \dot{\xi}(T)} \hat{U}_\xi (T,0) e^{-\frac{i}{\hbar} \hat{q} g \xi(0) \dot{\xi}(0)} | \psi_\omega \rangle\,.
\label{ifbasic2}
\ee
In this expression $F_{\psi_\omega}[\xi,\xi']$ does not depend on $\xi(t)$, $\xi'(t)$ for $t<0$ and $t>T$. Thus the path integrals over $\xi$ and $\xi'$ in~\eqref{nextprob} can be reduced to path integrals from $0$ to $T$ by introducing the freely-evolved wave functions
\ba
\tilde{\phi}_{A}\left(\tilde{\xi}_i\right)&=& \int d\xi_i\, \phi_A(\xi_i)\underset{\substack{\xi(-\Delta)=\xi_i\\\xi(0)=\tilde{\xi}_i}}\int\tilde{\mathcal{D}}\xi\, e^{\frac{i}{\hbar}\int_{-\Delta}^0dt\frac{1}{2}m_0\dot{\xi}^2}\ , \\
\tilde{\phi}_{B}\left(\tilde{\xi}_f\right)&=& \int d\xi_f\, \phi_B(\xi_f)\underset{\substack{\xi(T+\Delta)=\xi_f\\\xi(T)=\tilde{\xi}_f}}\int\tilde{\mathcal{D}}\xi\, e^{-\frac{i}{\hbar}\int_{T}^{T+\Delta}dt\frac{1}{2}m_0\dot{\xi}^2}\ ,
\ea
as well as their $\xi'$ counterparts. Dropping the tildes we can therefore write
\ba
P_{\psi_\omega}(\phi_A \to \phi_B)
&\equiv& \int d \xi_i d \xi'_i d \xi_f d \xi'_f \, \phi^*_A(\xi'_i)  \phi_B(\xi'_f) \phi^*_B(\xi_f) \phi_A(\xi_i)\nn\\
&&\times \underset{\substack{\xi(0) = \xi_i \; , \; \xi'(0) = \xi'_i \\ \xi(T) = \xi_f \; , \; \xi'(T) = \xi'_f}} \int \hspace{-10mm} \tilde{\cal D}\xi \tilde{\cal D}\xi' e^{\frac{i}{\hbar} \int_{0}^{T} dt \frac{1}{2} m_0 (\dot{\xi}^2 - \dot{\xi}'^{2})} F_{\psi_\omega}[\xi,\xi']\ , 
\label{transfinal}
\ea
and we see that the arbitrary time scale $\Delta$ has disappeared from the expression; this is now a path integral from $0$ to $T$. At the expense of introducing additional ordinary integrals, we can also assume that the values of $\dot{\xi}$, $\ddot{\xi}$ and $\dot{\xi}'$, $\ddot{\xi}'$ are fixed at $t=0$ and $t=T$, but to reduce clutter we do not make this explicit in our formulas. Putting everything together, we  see that the influence functional now depends explicitly on the boundary conditions in the path integral:
\be
F_{\psi_\omega}[\xi, \xi'] = \langle \psi_\omega | e^{\frac{i}{\hbar} \hat{q} g \xi'_i \dot{\xi}'_i} \hat{U}_{\xi'}^\dagger (T, 0)e^{-\frac{i}{\hbar} \hat{q} g \xi'_f \dot{\xi}'_f} e^{\frac{i}{\hbar} \hat{q} g \xi_f \dot{\xi}_f}  \hat{U}_\xi (T,0) e^{-\frac{i}{\hbar} \hat{q} g \xi_i \dot{\xi}_i} | \psi_\omega \rangle\ .
\ee
Our goal is to evaluate this for different harmonic oscillator states, but before we do that we can manipulate this expression further.

Let us split the Hamiltonian, \eqref{Hxi}, into a time-independent free piece and an interaction piece, $\hat{H}_{\xi} = \hat{H}_0 + \hat{H}_{\rm int}[\xi]$, where
\ba
\hat{H}_0 &\equiv& \frac{\hat{p}^2}{2m} + \frac{1}{2} m \omega^2 \hat{q}^2\ , \\
\hat{H}_{\rm int}[\xi] &\equiv& \frac{g\hat{p}\xi\dot\xi}{m} + \frac{g^2 \xi^2\dot{\xi}^2}{2m}\ .
\ea
Then the influence functional becomes
\be
F_{\psi_\omega}[\xi, \xi'] = \langle \psi_\omega | e^{\frac{i}{\hbar} \hat{q} g \xi'_i \dot{\xi}'_i} \hat{U}^{\rm int}_{\xi'}{}^\dagger (T)e^{-\frac{i}{\hbar} \hat{q}_I(T) g \xi'_f \dot{\xi}'_f} e^{\frac{i}{\hbar} \hat{q}_I(T) g \xi_f \dot{\xi}_f} \hat{U}^{\rm int}_\xi (T) e^{-\frac{i}{\hbar} \hat{q} g \xi_i \dot{\xi}_i} | \psi_\omega \rangle\ ,
\label{firinf}
\ee
where quantities with a label $I$ are understood to be in the interaction picture ({\it e.g.} $\hat{q}_I(t)=e^{i\hat{H}_0t/\hbar}\hat{q}e^{-i\hat{H}_0t/\hbar}$) and
\be
\hat{U}^{\rm int}_\xi(T) \equiv {\cal T} \left ( e^{- \frac{i}{\hbar} \int_0^T \hat{H}^I_{\rm int}[\xi] dt} \right )\,,
\ee
is the interaction-picture time-evolution operator, expressed as a time-ordered exponential. Since in the interaction picture, $\hat{p}_I=m\dot{\hat{q}}_I$, we can write the interaction Hamiltonian as
\be
\hat{H}_{\rm int}^I[\xi]=g\dot{\hat{q}}_I\xi\dot\xi + \frac{g^2 \xi^2\dot{\xi}^2}{2m}\ . 
\ee
Then the commutator $\left[\hat{H}_{\rm int}^I[\xi(t)], \hat{H}_{\rm int}^I[\xi(t')]\right]=g^2\xi(t)\dot{\xi}(t)\xi(t')\dot{\xi}(t')\left[\dot{\hat{q}}_I(t),\dot{\hat{q}}_I(t')\right]$ is seen to be a $c$-number (as are any commutators involving only the operators $\hat{q}_I$ and $\hat{p}_I=m\dot{\hat{q}}_I$). Consequently we can eliminate the time-ordering symbol at the expense of an additional term in the exponent~\cite{Itzykson:1980rh}:
\ba
\hat{U}^{\rm int}_\xi(T) &=&\exp\left(- \frac{i}{\hbar} \int_0^T \hat{H}^I_{\rm int}[\xi] dt-\frac{1}{2\hbar^2}\int_0^T\int_0^t dt dt'\left[\hat{H}_{\rm int}^I[\xi(t)], \hat{H}_{\rm int}^I[\xi(t')]\right]\right)\nn\\
&=&\exp\left(- \frac{ig}{\hbar} \int_0^T \dot{\hat{q}}_I(t)\xi(t)\dot\xi(t) dt\right)\nn\\
&&\hspace{-4mm}\times
\exp\left(- \frac{ig^2}{2m\hbar} \int_0^T \xi^2(t)\dot\xi^2(t) dt-\frac{g^2}{2\hbar^2}\int_0^T\int_0^t dt dt'\,\xi(t)\dot{\xi}(t)\xi(t')\dot{\xi}(t')\left[\dot{\hat{q}}_I(t),\dot{\hat{q}}_I(t')\right]\right).
\ea
After repeated use of integration by parts to remove the time derivatives from the $\hat{q}_I$ operators this expression becomes
\ba
\hat{U}^{\rm int}_\xi(T) &=& \exp\left(\frac{ig}{2\hbar} \int_0^T dt\,\hat{q}_I(t)X(t) -\frac{ig}{\hbar}\hat{q}_I(T)\xi_f\dot\xi_f+\frac{ig}{\hbar}\hat{q}\xi_i\dot\xi_i\right)\nn\\
&&\times\exp\Biggr(-\frac{g^2}{8\hbar^2}\int_0^T\int_0^t dt dt'\,\left[\hat{q}_I(t),\hat{q}_I(t')\right]X(t)X(t')-\frac{g^2}{4\hbar^2}\int_0^Tdt\left[\hat{q}_I(t),\hat{q}\right]\xi_i\dot\xi_iX(t)\nn\\
&&\hspace{13mm}+\frac{g^2}{4\hbar^2}\int_0^Tdt'\left[\hat{q}_I(T),\hat{q}_I(t')\right]\xi_f\dot\xi_fX(t')+\frac{g^2}{2\hbar^2}\left[\hat{q}_I(T),\hat{q}\right]\xi_i\dot\xi_i\xi_f\dot\xi_f\Biggr)\ ,
\label{bigeuxi}
\ea
where $\hat{q} = \hat{q}_I(0)$. Here, to avoid writing cumbersome second derivatives of $\xi^2$, we have introduced
\ba
X(t) & \equiv & \frac{d^2}{dt^2} \xi^2(t)\,,\\
X'(t) & \equiv & \frac{d^2}{dt^2} \xi'^2(t)\,,
\label{Xintermsofxi}
\ea
the latter definition being included for later convenience.  

Next we invoke the relation
\be
e^{\hat{A}} e^{\hat{B}} = e^{\hat{A}+\hat{B}} e^{\frac{1}{2} [\hat{A},\hat{B}]}\,,
\label{bch}
\ee
a variant of the Baker-Campbell-Hausdorff formula valid when $[\hat{A},\hat{B}]$ is a $c$-number. This formula allows us to reduce~\eqref{bigeuxi}:
\be
\hat{U}^{\rm int}_\xi(T) = e^{-\frac{ig}{\hbar}\hat{q}_I(T)\xi_f\dot\xi_f}e^{\frac{ig}{2\hbar} \int_0^T dt\,\hat{q}_I(t)X(t)}e^{\frac{ig}{\hbar}\hat{q}\xi_i\dot\xi_i} e^{-\frac{g^2}{8\hbar^2}\int_0^T\int_0^t dt dt'\,\left[\hat{q}_I(t),\hat{q}_I(t')\right]X(t)X(t')}\ .
\ee
With this expression and its $\xi'$ counterpart at hand, we can dramatically simplify the form of the influence functional~\eqref{firinf}. We find
\be
F_{\psi_\omega}[\xi,\xi']=e^\mathcal{S}\langle\psi_\omega| e^{-\frac{ig}{2\hbar}\int_0^T dt \hat{q}_I(t)X'(t)} e^{\frac{ig}{2\hbar}\int_0^T dt \hat{q}_I(t)X(t)} |\psi_\omega\rangle\,,
\label{interinf}
\ee
where 
\be
\mathcal{S}\equiv\frac{g^2}{8\hbar^2}\int_0^T\int_0^t dt\,dt'[\hat{q}_I(t),\hat{q}_I(t')]\left(X'(t)X'(t')-X(t)X(t')\right)\,.
\ee

Further simplification can be achieved by defining the ladder operators
$\hat{a}$ and $\hat{a}^\dag$ in the usual way:
\ba
\hat{a}\equiv\sqrt{\frac{m\omega}{2\hbar}}\left(\hat{q}+\frac{i}{m\omega}\hat{p}\right)\,,\\
\hat{a}^\dag\equiv\sqrt{\frac{m\omega}{2\hbar}}\left(\hat{q}-\frac{i}{m\omega}\hat{p}\right)\,.
\ea
Then
\be
\hat{q}_I(t)=\sqrt{\frac{\hbar}{2m\omega}}\left(\hat{a}e^{-\omega t}+\hat{a}^\dagger e^{i\omega t}\right)\ ,
\ee
and we can repeatedly invoke~\eqref{bch} to bring the matrix element in~\eqref{interinf} into normal order. We arrive, finally, at a suitable form of the influence functional: 
\be
F_{\psi_\omega}[\xi,\xi']
= F_{0_\omega}[\xi,\xi'] \langle \psi_\omega | e^{-W^* \hat{a}^\dagger} e^{W \hat{a}} | \psi_\omega \rangle \ .
\label{geninffunc}
\ee
Here
\be
W \equiv \frac{ig}{\sqrt{8m\hbar\omega}}\int_0^T dt\left(X(t)-X'(t)\right)e^{-i\omega t}\,,	\label{W}
\ee
and
\be
F_{0_\omega}[\xi,\xi']\equiv \exp\left[-\frac{g^2}{8m\hbar\omega}\int_0^T\int_0^t dt\,dt'\left(X(t)-X'(t)\right)\left(X(t')e^{-i\omega(t-t')}-X'(t')e^{i\omega(t-t')}\right)\right]\ , \label{vacinffunc}
\ee
where $X(t) = \frac{d^2}{dt^2} \xi^2(t)$ and $X'(t) = \frac{d^2}{dt^2} \xi'^2(t)$. Evidently $F_{0_\omega}[\xi,\xi']$ is the influence functional of the ground state, as can be seen from~\eqref{geninffunc} when $|\psi_\omega\rangle=|0_\omega\rangle$. For future reference, we note the influence phase of the ground state:
\be
i\Phi_{0_\omega}[\xi,\xi']=-\frac{g^2}{8m\hbar\omega}\int_0^T\int_0^t dt\,dt'\left(X(t)-X'(t)\right)\left(X(t')e^{-i\omega(t-t')}-X'(t')e^{i\omega(t-t')}\right)\ .
\label{vacinfphase}
\ee
We can now in principle compute the influence functional for arbitrary states $|\psi_\omega\rangle$ of the incoming gravitational field mode. However, we cannot yet evaluate the ground state contribution $F_{0_\omega}[\xi, \xi']$ itself because it depends on the unphysical mass $m$, which in turn depends on the infrared regulator $L$ that we used in our finite-volume discretization of the modes. (Actually, $m$ also appears in $W$, but this dependence sometimes drops out.) We will sort this out in Section~\ref{qmfield} when we sum over modes.

\subsection*{Example: Coherent States}

\noindent
As an illustrative example, consider a graviton mode of energy $\hbar\omega$ in a coherent state: $|\psi_\omega \rangle = |\alpha_\omega \rangle$. Here $\alpha_\omega$ is the eigenvalue of the annihilation operator, $\hat{a}$:
\be
\hat{a}|\alpha_\omega\rangle=\alpha_\omega|\alpha_\omega\rangle\ .	\label{cohalpha}
\ee
Since $\hat{a}$ is not hermitian, $\alpha_\omega$ can be a complex number. Physically, coherent states are the quantum states that most closely resemble solutions of the classical equations of motion. Consider a classical gravitational wave mode: 
\be
q_{\rm cl}(t)\equiv Q_\omega\cos(\omega t +\varphi_\omega)\ .
\ee
We can find the corresponding value of $\alpha_\omega$ by noting that
\ba
\langle\alpha_\omega|\hat{q}|\alpha_\omega\rangle&=&q_{\rm cl}(t=0)=Q_\omega\cos\varphi_\omega\ ,\\
\langle\alpha_\omega|\hat{p}|\alpha_\omega\rangle&=&m\dot{q}_{\rm cl}(t=0)=-m\omega Q_\omega\sin\varphi_\omega\ .
\label{cohclassical}
\ea
Hence
\be
\alpha_\omega =\sqrt{\frac{m\omega}{2\hbar}}Q_\omega e^{-i\varphi_\omega}\,.
\ee
Let us now calculate the influence functional in the state $|\alpha_\omega \rangle$. From (\ref{geninffunc}) and (\ref{cohalpha}), we see immediately that
\be
F_{\alpha_\omega}[\xi,\xi']
= F_{0_\omega}[\xi,\xi']e^{-W^* \alpha_\omega^*+W\alpha_\omega}\,.
\ee
Substituting (\ref{W}), we find
\be
F_{\alpha_\omega}[\xi,\xi']=F_{0_\omega}[\xi,\xi']\exp\left[\frac{ig}{2\hbar}\int_0^T dt\,Q_\omega\cos(\omega t+\varphi_\omega)\left(X(t)-X'(t)\right)\right]\,. 
\label{Fcoh}
\ee
We have thus calculated the influence functional for a mode in a coherent state, up to evaluation of the ground state influence functional, $F_{0_\omega}[\xi,\xi']$. Inserting this expression into the transition probability, \eqref{transfinal}, we find
\ba
P_{\alpha_\omega}(\phi_A \to \phi_B)
&\equiv& \int d \xi_i d \xi'_i d \xi_f d \xi'_f \, \phi^*_A(\xi'_i)  \phi_B(\xi'_f) \phi^*_B(\xi_f) \phi_A(\xi_i)\times\nn\\
&&\hspace{-40mm}\underset{\substack{\xi(0) = \xi_i \; , \; \xi'(0) = \xi'_i \\ \xi(T) = \xi_f \; , \; \xi'(T) = \xi'_f}} \int \hspace{-10mm} \tilde{\cal D}\xi \tilde{\cal D}\xi' \exp\left[\frac{i}{\hbar} \int_{0}^{T} dt \left\{\frac{1}{2} m_0 \left(\dot{\xi}^2 - \dot{\xi}'^{2}\right)+\frac{1}{2}gQ_\omega\cos(\omega t+\varphi_\omega)\left(X(t)-X'(t)\right)\right\}\right] F_{0_\omega}[\xi,\xi']\ . \label{Pcoh}
\ea
Let us interpret this result. We see that when the detector encounters a quantized gravitational wave mode -- a coherent state -- its transition probability is affected in two ways. There is, as always, the ground state influence functional $F_{0_\omega}[\xi,\xi']$. In addition, the Lagrangian picks up a piece $\frac{1}{2}gQ_\omega \cos(\omega t +\phi_\omega)\frac{d^2}{d\xi^2}\xi^2(t)$. But observe that, after an integration by parts, this is precisely the interaction Lagrangian in~\eqref{Somega} with $q=q_{\rm cl}$. In other words, the dynamics of the detector arm is merely modified to incorporate the background classical gravitational wave; the only effect with a purely quantum origin is the ground state fluctuation encoded in $F_{0_\omega}[\xi,\xi']$, which would have been present even in the absence of the coherent state. Put another way, there is no way to discern the gravitons that specifically comprise a classical gravitational wave.

More generally, one can ``add'' a classical configuration to any other state vector $|\chi_\omega\rangle$ through the action of the unitary displacement operator
\be
\hat{D}(\alpha_\omega)\equiv e^{\alpha_\omega\hat{a}^\dag-\alpha_\omega^*\hat{a}}\ .
\label{displacement}
\ee
Suppose then that $|\psi_\omega\rangle = \hat{D}(\alpha_\omega)|\chi_\omega\rangle$. This generalizes our earlier coherent state $|\alpha_\omega\rangle$ which could have been written as $\hat{D}(\alpha_\omega)|0_\omega\rangle$. The displacement operator has the properties
\ba
\hat{D}(\alpha_\omega)^\dag\hat{a}\hat{D}(\alpha_\omega)&=&\hat{a}+\alpha_\omega\,,\\
\hat{D}(\alpha_\omega)^\dag\hat{a}^\dagger \hat{D}(\alpha_\omega)&=&\hat{a}^\dag+\alpha_\omega^*\,.
\ea
Then the corresponding influence functional is
\ba
F_{\psi_\omega}[\xi,\xi']&=&F_{0_\omega}[\xi,\xi'] \langle \chi_\omega |\hat{D}(\alpha_\omega)^\dag e^{-W^* \hat{a}^\dagger} e^{W \hat{a}} \hat{D}(\alpha_\omega)|  \chi_\omega \rangle \nn\\
&=&F_{0_\omega}[\xi,\xi'] \langle  \chi_\omega | e^{-W^* \hat{D}(\alpha_\omega)^\dag\hat{a}^\dagger \hat{D}(\alpha_\omega)} e^{W \hat{D}(\alpha_\omega)^\dag\hat{a}\hat{D}(\alpha_\omega)}|  \chi_\omega \rangle\nn\\
&=&F_{\chi_\omega}[\xi,\xi'] e^{-W^* \alpha_\omega^*+W\alpha_\omega} \nn\\
&=&F_{\chi_\omega}[\xi,\xi'] \exp\left[\frac{ig}{2\hbar}\int_0^T dt\,Q_\omega\cos(\omega t+\varphi_\omega)\left(X(t)-X'(t)\right)\right]\,.
\ea
As before, the overall effect of a displacement operator is simply to modify the classical action; any intrinsically quantum contributions to the influence functional must originate from the state $|\chi_\omega\rangle$.

\section{Quantized Gravitational Field Coupled to the Detector}
\label{qmfield}

\noindent
Having computed the influence functional for a single gravitational field mode, we are now ready to tackle the general problem of a continuum of modes -- a quantum field -- interacting with the detector. The quantum state of the gravitational field $|\Psi\rangle$ can be written as a tensor product of the Hilbert states of the individual graviton modes:
\be
|\Psi \rangle = \bigotimes_{\vec{k}}|\psi_{\omega(\vec{k})}\rangle\ .
\ee
Since the action for the field, (\ref{genaction}), involves a sum over modes, the field influence functional is a product of the mode influence functionals:
\be
F_\Psi[\xi,\xi'] = \prod_{\vec{k}} F_{\psi_{\omega(\vec{k})}} [\xi, \xi']\ .
\ee
Correspondingly, the field influence phase is a sum over the influence phases for each mode:
\be
\Phi_\Psi[\xi,\xi'] = \sum_{\vec{k}} \Phi_{\psi_{\omega(\vec{k})}} [\xi, \xi']\ .
\ee
Note that when summing over modes our choice of the mode action~\eqref{Somega} (motivated by simplicity) breaks down in a number of ways. For a given arm orientation, the cross ($\times$) polarization cannot be neglected for all $\vec{k}$. Moreover, a  mode with a generic wave vector $\vec{k}$ will excite all three degrees of freedom of the detector arm~\eqref{genaction}. Lastly, a more careful treatment of the spatial integration over modes with wave vectors non-parallel to the $z$-axis will yield additional trigonometric factors of order one. We leave all such refinements to future work. In the rest of this section, we evaluate this mode sum for different field states. This will allow us in Section~\ref{langevin} to determine the quantum-influenced dynamics of the arm length.

\subsection{Vacuum state}

\noindent
When the gravitational field is in its vacuum state, $|\Psi\rangle=|0\rangle$, all the modes are in their corresponding ground states. The vacuum influence function
\be
F_0[\xi,\xi']=e^{i\Phi_0[\xi,\xi']}\ ,
\ee 
can therefore be written as a product of the ground state influence functionals. Correspondingly, the vacuum influence phase is a mode sum over the ground state influence phases~\eqref{vacinfphase}:
\ba
i\Phi_0[\xi,\xi']
&=& \sum_{\vec{k}}i\Phi_{0_{\omega(\vec{k})}}[\xi,\xi']\nn\\
 &=&-\sum_{\vec{k}}\frac{g^2}{8m\hbar\omega}\int_0^T\int_0^t dt\,dt'\left(X(t)-X'(t)\right)\left(X(t')e^{-i\omega(t-t')}-X'(t')e^{i\omega(t-t')}\right)\nn\\
 &=&-\frac{m_0^2 G}{16\pi^2\hbar}\frac{(2\pi)^3}{L^3}\sum_{\vec{k}}\frac{1}{\omega}\int_0^T\int_0^t dt\,dt'\left(X(t)-X'(t)\right)\left(X(t')e^{-i\omega(t-t')}-X'(t')e^{i\omega(t-t')}\right)\nn\\
 &=& -\frac{m_0^2 G}{4\pi\hbar}\int_0^\infty\omega d\omega\int_0^T\int_0^t dt\,dt'\left(X(t)-X'(t)\right)\left(X(t')e^{-i\omega(t-t')}-X'(t')e^{i\omega(t-t')}\right)\nn\\
 &=&-\frac{m_0^2 G}{4\pi\hbar}\int_0^T\int_0^t dt\,dt' \int_0^\infty d\omega\,\omega\cos(\omega(t-t'))\,\left(X(t)-X'(t)\right)\left(X(t')-X'(t')\right)\nn\\
&&+\frac{im_0^2 G}{4\pi\hbar}\int_0^T\int_0^t dt\,dt' \int_0^\infty d\omega\,\omega\sin(\omega(t-t'))\left(X(t)-X'(t)\right)\left(X(t')+X'(t')\right)\ .
\label{ifvaclong}
\ea
Here we have taken the continuum limit of the mode sum and replaced $m$ and $g$ by their values in terms of physical constants via \eqref{mdef} and \eqref{gdef}; the unphysical volume of space $L^3$ has thereby dropped out. 
 
Notice, however, that the $\omega$ integrals are divergent. Nevertheless, as we shall see in Section~\ref{langevin}, this expression enables us to calculate physically meaningful (and finite) effects on the dynamics of the arm length. In particular, the real and imaginary parts of the last line of~\eqref{ifvaclong} will have an interpretation, in the context of the fluctuation-dissipation theorem, as Gaussian noise and radiation loss.

\subsection{Coherent states}

\noindent
In quantum field theory, coherent states are the quantum states that most closely resemble classical solutions of the field equations. Specifically, the expectation value of a field operator in a coherent state is precisely the classical value of the field; indeed, we used just this property when we considered coherent states in a quantum-mechanical context, \eqref{cohclassical}. These states arise naturally when dealing with classical gravitational waves: since signal templates are obtained by solving Einstein's equations, they all correspond quantum-mechanically to coherent states of the gravitational field.

Consider, then, a gravitational plane wave propagating along the $z$ axis with $+$ polarization. Its wave profile, also known as the strain, can be written as
 \be
 \bar{h}(t)\equiv \frac{1}{\sqrt{\hbar G}}\sum_{\omega} Q_\omega \cos(\omega t+\varphi_\omega)\ .
 \label{waveprofile}
 \ee
Of course gravitational waves emitted by a localized source situated at a finite distance $r$ are more appropriately described by spherical waves. For sufficiently distant sources, however, the plane wave approximation is excellent and the appropriate $1/r$ decay factor is built into the amplitude $Q_\omega$. As seen in Section~\ref{qmdet}, each of the modes in~\eqref{waveprofile} is described by a quantum-mechanical coherent state $|\alpha_{\omega}\rangle$ with 
\be
\alpha_\omega=\sqrt{\frac{m\omega}{2\hbar}}\,Q_\omega e^{-i\varphi_\omega}\ .
\ee 
Thus the field-theoretic coherent state corresponding to $\bar{h}$ is given by
\be
|\bar{h}\rangle = \bigotimes_{\omega}|\alpha_{\omega}\rangle\ .
\ee
When the gravitational field is in this state, $|\Psi\rangle=|\bar{h}\rangle$, the  influence functional
\be
F_{\bar{h}}[\xi,\xi']=F_0[\xi,\xi']e^{i\Phi_{\bar{h}}[\xi,\xi']}\ ,
\ee 
is a product of the quantum-mechanical coherent state influence functionals, \eqref{Fcoh}, for wave vectors parallel to the $z$-axis, and a product of ground state influence functionals for all other wave vectors. For the coherent part of the influence phase we then have
\be
i\Phi_{\bar{h}}[\xi,\xi'] =\frac{i}{\hbar}\int_0^T dt\frac{1}{4}m_0\bar{h}(t)\left(X(t)-X'(t)\right)\ ,
\ee
and the total influence phase is $\Phi_0+\Phi_{\bar{h}}$. The transition probability between states $A$ and $B$ of the detector is therefore
 \ba
P_{\bar{h}}(\phi_A \to \phi_B)
&\equiv& \int d \xi_i d \xi'_i d \xi_f d \xi'_f \, \phi^*_A(\xi'_i)  \phi_B(\xi'_f) \phi^*_B(\xi_f) \phi_A(\xi_i)\times\nn\\
&&\hspace{-30mm}\underset{\substack{\xi(0) = \xi_i \; , \; \xi'(0) = \xi'_i \\ \xi(T) = \xi_f \; , \; \xi'(T) = \xi'_f}} \int \hspace{-10mm} \tilde{\cal D}\xi \tilde{\cal D}\xi' \exp\left[\frac{i}{\hbar} \int_{0}^{T} dt \left\{\frac{1}{2} m_0 \left(\dot{\xi}^2 - \dot{\xi}'^{2}\right)+\frac{1}{4}m_0\bar{h}(t)\left(X(t)-X'(t)\right)\right\}\right] F_0[\xi,\xi']\ .
\label{cohtrans}
\ea
As we saw with individual graviton modes in coherent states, \eqref{Pcoh}, the only effect on a detector interacting with a quantized gravitational wave -- besides the omnipresent vacuum fluctuations encoded in $F_0$ -- is to contribute to the action a piece that corresponds to an interaction with a classical gravitational wave, $\bar{h}(t)$. Although one might perhaps have expected quantum effects akin to graviton shot noise, we see that (other than vacuum fluctuations) there is no specific signature of the quantization of  gravitational waves emitted by a classical source.

\subsection{Thermal states}

\noindent
Now let us consider a slightly different example for which the gravitational field is in a thermal state. In practice this could describe a cosmic gravitational wave background: although each gravitational wave is described by a coherent state, their incoherent superposition is not. Alternatively, a thermal gravitational field state can also be sourced by an evaporating black hole.

Thermal states are mixed states and as such are described by density matrices; the extension of the influence functional formalism to this setting is straightforward. For a single-mode density matrix $\rho_\omega$, the generalization of~\eqref{geninffunc} is 
\be
F_{\rho_\omega}[\xi,\xi']
= F_{0_\omega}[\xi,\xi'] {\rm Tr}\left[\rho_\omega\, e^{-W^* \hat{a}^\dagger} e^{W \hat{a}}\right]\ ,
\ee
where $W$ is given by~\eqref{W}. For a thermal state at temperature $T$, the density matrix for a mode of energy $\hbar\omega$ is
\be
\rho^{\rm th}_\omega=\frac{1}{Z}\sum_{n=0}^\infty e^{-\frac{\hbar\omega(n+1/2)}{k_BT}}|n\rangle\langle n|\ ,
\ee
where $Z=\sum_{n=0}^{\infty}\exp\left(-\frac{\hbar\omega(n+1/2)}{k_BT}\right)$ is the partition function, $k_B$ is Boltzmann's constant, and $|n\rangle$ is an energy eigenstate. Then the quantum-mechanical single-mode influence functional reads
\ba
F^{\rm th}_\omega[\xi,\xi']
&=&\left(1-e^{-\frac{\hbar\omega}{k_BT}}\right)F_{0_\omega}[\xi,\xi'] \sum_{n=0}^\infty e^{-\frac{\hbar\omega n}{k_BT}}\langle n|e^{-W^* \hat{a}^\dagger} e^{W \hat{a}}|n\rangle\nn\\
&=& \left(1-e^{-\frac{\hbar\omega}{k_BT}}\right)F_{0_\omega}[\xi,\xi'] \sum_{n=0}^\infty\sum_{p=0}^\infty\sum_{q=0}^\infty \frac{(-1)^pW^*{}^pW^q}{p!q!}e^{-\frac{\hbar\omega n}{k_BT}}\langle n|\hat{a}^\dagger{}^p \hat{a}^q|n\rangle\nn\\
&=& \left(1-e^{-\frac{\hbar\omega}{k_BT}}\right)F_{0_\omega}[\xi,\xi'] \sum_{p=0}^\infty \frac{(-1)^p|W|^{2p}}{p!^2}e^{-\frac{\hbar\omega p}{k_BT}}\sum_{n=p}^\infty n(n-1)\dots (n-p+1)e^{-\frac{\hbar\omega (n-p)}{k_BT}}\nn\\
&=& \left(1-e^{-\frac{\hbar\omega}{k_BT}}\right)F_{0_\omega}[\xi,\xi'] \sum_{p=0}^\infty \frac{(-1)^p|W|^{2p}}{p!^2}e^{-\frac{\hbar\omega p}{k_BT}}\frac{d^p}{dx^p}\left[\frac{1}{1-x}\right]_{x=e^{-\frac{\hbar\omega}{k_BT}}}\nn\\
&=& F_{0_\omega}[\xi,\xi'] \sum_{p=0}^\infty \frac{(-1)^p|W|^{2p}}{p!}\left(\frac{e^{-\frac{\hbar\omega }{k_BT}}}{1-e^{-\frac{\hbar\omega}{k_BT}}}\right)^p\nn\\
&=&F_{0_\omega}[\xi,\xi']\exp\left[-|W|^2\left(\frac{1}{e^{\frac{\hbar\omega}{k_BT}}-1}\right)\right]\,.
\label{Fth}
\ea
We see that the thermal influence functional features an exponential factor multiplying the ground state influence functional. 

Now let us extend this result to a thermal gravitational field state. The thermal field density matrix is a tensor product of the mode density matrices:
\be
\rho_{\rm th}=  \bigotimes_{\vec{k}}\rho^{\rm th}_{\omega(\vec{k})}\ .
\ee
We can sum over all modes to obtain
\be
F_{\rm th}[\xi,\xi']=F_0[\xi,\xi']e^{i\Phi_{\rm th}[\xi,\xi']}\, ,
\ee
where, using~\eqref{Fth}, we have
\ba
i\Phi_{\rm th}[\xi,\xi']&=&-\frac{m_0^2 G}{4\pi\hbar}\int_0^\infty \frac{\omega d\omega}{e^{\frac{\hbar\omega}{k_BT}}-1} \int_0^T\int_0^T dt\,dt'\left(X(t)-X'(t)\right)\left(X(t')-X'(t')\right)e^{-i\omega(t-t')}\nn\\
&=&-\frac{m_0^2 G}{4\pi\hbar}\int_0^\infty \frac{\omega d\omega}{e^{\frac{\hbar\omega}{k_BT}}-1} \int_0^T\int_0^T dt\,dt'\left(X(t)-X'(t)\right)\left(X(t')-X'(t')\right)\cos(\omega(t-t'))\ .
\label{phithermal}
\ea
Symmetry under $t\leftrightarrow t'$ ensures that the sine part of the complex exponential does not contribute; that the result is real can also be seen from~\eqref{Fth}. Note that in performing the mode sum we have integrated over all wave vectors $\vec{k}$; this would be appropriate for an isotropic cosmic background. However, for a localized, evaporating black hole, the state is thermal only for those wave vectors $\vec{k}$ that point within the solid angle subtended by the black hole. This would result in the thermal part of the influence phase being multiplied by a minuscule factor of $\frac{1}{2}(1-\cos\theta_0) \sim \frac{1}{4}\left(\frac{r_S}{r}\right)^2$ where $\theta_0$ is the half-angle subtended by the black hole, $r_S$ is its Schwarzschild radius, and $r \gg r_S$ its distance from the detector.

\subsection{Squeezed Vacua}

\noindent
So far we have considered quantum states of the gravitational field that have a straightforward classical interpretation. We will now examine squeezed states which exhibit more distinctly quantum-mechanical features. Physically, such states are conjectured to arise in post-inflationary scenarios~\cite{Grishchuk:1990bj,Albrecht:1992kf}. In quantum mechanics, squeezed states have the characteristic property that uncertainties in certain operators, say $\hat{q}$ or $\hat{p}$, are smaller than $\hbar/2$. They are constructed with the help of the unitary squeezing operator
\be
\hat{S}(z)\equiv e^{\frac{1}{2}(z^*\hat{a}^2+z\hat{a}^\dag{}^2)}\,,
\ee
where $z$ is a complex number known as the squeezing parameter. 
A squeezed ground state for instance is $\hat{S}(z)|0\rangle$ and one can also define squeezed coherent states, $\hat{S}(z)\hat{D}(\alpha)|0\rangle$, which combine the squeezing operator with the displacement operator~\eqref{displacement}. 

Let us consider the gravitational field to be in a squeezed vacuum, for which each mode of energy $\hbar\omega$ is in a squeezed ground state $\hat{S}(z_\omega)|0_\omega\rangle$. Then the single-mode influence functional is
\ba
F_{z_\omega}[\xi,\xi']&=&F_{0_\omega}[\xi,\xi'] \langle 0_\omega |\hat{S}(z_\omega)^\dag e^{-W^* \hat{a}^\dagger} e^{W \hat{a}} \hat{S}(z_\omega)| 0_\omega \rangle \nn\\
&=&F_{0_\omega}[\xi,\xi'] \langle 0_\omega | e^{-W^* \hat{S}(z_\omega)^\dag\hat{a}^\dagger \hat{S}(z_\omega)} e^{W \hat{S}(z_\omega)^\dag\hat{a}\hat{S}(z_\omega)}| 0_\omega \rangle\nn\\
&=&F_{0_\omega}[\xi,\xi'] \exp\left[-\frac{1}{4}\left(W^*{}^2 e^{-i\phi_\omega}+W^2 e^{i\phi_\omega}\right)\sinh 2r_\omega-\frac{1}{2}|W|^2(\cosh 2r_\omega -1)\right]\ .
\label{squeezed}
\ea
Here we have defined $z_\omega\equiv r_\omega e^{i\phi_\omega}$ and we have invoked~\eqref{bch} as well as 
\ba
S(z_\omega)^\dag\hat{a}S(z_\omega)&=&\cosh r_\omega\, \hat{a}-e^{i\phi_\omega}\sinh r_\omega\,\hat{a}^\dag\ ,\\
S(z_\omega)^\dag\hat{a}^\dagger S(z_\omega)&=&\cosh r_\omega\, \hat{a}^\dag-e^{-i\phi_\omega}\sinh r_\omega\, \hat{a}\ .
\ea
We can rewrite \eqref{squeezed} as $F_{z_\omega}[\xi,\xi']=F_{0_\omega}[\xi,\xi']e^{i\Phi_{z_\omega}[\xi,\xi']}$,
where
\ba
i\Phi_{z_\omega}[\xi,\xi'] &=&-\frac{g^2}{16m\hbar\omega}\int_0^T\int_0^T dt\,dt'\left(X(t)-X'(t)\right)\left(X(t')-X'(t')\right)\cos(\omega(t-t'))(\cosh 2r_\omega-1)\nn\\
&&+\frac{g^2}{16m\hbar\omega}\int_0^T\int_0^T dt\,dt'\left(X(t)-X'(t)\right)\left(X(t')-X'(t')\right)\cos(\omega(t+t')-\phi_\omega)\sinh 2r_\omega\ .
\ea
Before we can sum over all modes we need to specify the amount of squeezing per mode $z_\omega$. An analysis of realistic squeezing parameters is beyond the scope of the current work; for the sake of simplicity, we will choose $r_\omega$ to be independent of $\omega$ and $\phi_\omega$ to be zero. Summing over all modes then yields the field-theoretic influence functional
\be
F_{z}[\xi,\xi']=F_{0}[\xi,\xi']e^{i\Phi_{z}[\xi,\xi']}\ ,
\ee
where
\ba
i\Phi_{z}[\xi,\xi'] &=&-\frac{m_0^2G}{8\pi\hbar}(\cosh 2r-1)\int_0^T\int_0^Tdt\,dt'\int_0^\infty d\omega\,\omega \cos(\omega(t-t'))\left(X(t)-X'(t)\right)\left(X(t')-X'(t')\right)\nn\\
&&\hspace{-1cm}+\frac{m_0^2G}{8\pi\hbar}\sinh 2r\int_0^T\int_0^T dt\,dt'\int_0^\infty d\omega\,\omega\cos(\omega(t+t'))\left(X(t)-X'(t)\right)\left(X(t')-X'(t')\right)\ .
\label{phisqueezed}
\ea
Notice that the first term in this expression is proportional to the real part of $i\Phi_0[\xi,\xi']$, as seen from~\eqref{ifvaclong}. The second term breaks the time-translation symmetry $t\rightarrow t+\delta$, $t'\rightarrow t'+\delta$. We will analyze the effects of these properties in the following section.

\section{Effective Equation of Motion of the Detector}
\label{langevin}

\noindent
Let us now use our results to derive an effective, quantum-corrected equation of motion for the arm length $\xi$. The equation of motion in the presence of a purely classical gravitational perturbation is the Euler-Lagrange equation, which follows from the classical action:
\be
\ddot{\xi} - \frac{1}{2} \ddot{\bar{h}} \xi = 0\ .
\label{classeom}
\ee
The source term here is the usual tidal acceleration in the presence of a gravitational perturbation. The question we are now finally in a position to address is: how does this equation change when the gravitational field is quantized? 

We know that the effect on $\xi$ is encoded in the Feynman-Vernon influence functional, which in the previous sections we have painstakingly evaluated for several classes of quantum  states of the gravitational field. The transition probability for the detector in the presence of a gravitational field state $|\Psi \rangle = \bigotimes_{\vec{k}}|\psi_{\omega(\vec{k})}\rangle$ is the natural extension of~\eqref{transfinal}:
\ba
P_{\Psi}(\phi_A \to \phi_B)
&=& \int d \xi_i d \xi'_i d \xi_f d \xi'_f \, \phi^*_A(\xi'_i)  \phi_B(\xi'_f) \phi^*_B(\xi_f) \phi_A(\xi_i)\nn\\
&&\times \underset{\substack{\xi(0) = \xi_i \; , \; \xi'(0) = \xi'_i \\ \xi(T) = \xi_f \; , \; \xi'(T) = \xi'_f}} \int \hspace{-10mm} \tilde{\cal D}\xi \tilde{\cal D}\xi' e^{\frac{i}{\hbar} \int_{0}^{T} dt \frac{1}{2} m_0 (\dot{\xi}^2 - \dot{\xi}'^{2})} F_{\Psi}[\xi,\xi']\ . 
\label{transfinalfield}
\ea
This equation is readily understood. The four ordinary integrals encode the initial and final states of $\xi$; however, as we are interested in the effective equation of motion for $\xi$ -- which will arise from taking a saddle point of the path integrals -- they will play no role. The double path integrals reflect the fact that we are calculating probabilities rather than probability amplitudes. The exponent is seen to be of the form $\frac{i}{\hbar}(S_0[\xi]-S_0[\xi'])$ where $S_0$ is the free particle action. Crucially, the gravitational field has been integrated out and its effect is now fully captured by the influence functional $F_{\Psi}[\xi,\xi']=e^{i\Phi_\Psi[\xi,\xi']}$.

To see how the equation of motion~\eqref{classeom} becomes modified, let us start by considering a gravitational field in a coherent state, $|\Psi\rangle =|\bar{h}\rangle$. Then the transition probability is given by~\eqref{cohtrans}:  
\ba
P_{\bar{h}}(\phi_A \to \phi_B)
&\equiv& \int d \xi_i d \xi'_i d \xi_f d \xi'_f \, \phi^*_A(\xi'_i)  \phi_B(\xi'_f) \phi^*_B(\xi_f) \phi_A(\xi_i)\,\times\nn\\
&&\int \tilde{\cal D}\xi \tilde{\cal D}\xi' \exp\Biggr[\frac{i}{\hbar} \int_{0}^{T} dt \left\{\frac{1}{2} m_0 \left(\dot{\xi}^2 - \dot{\xi}'^{2}\right)+\frac{1}{4}m_0\bar{h}(t)\left(X(t)-X'(t)\right)\right\}\nn\\
&&\hspace{-10mm}-\frac{m_0^2 G}{4\pi\hbar}\int_0^T\int_0^t dt\,dt' \int_0^\infty d\omega\,\omega\cos(\omega(t-t'))\,\left(X(t)-X'(t)\right)\left(X(t')-X'(t')\right)\nn\\
&&\hspace{-10mm}+\frac{im_0^2 G}{4\pi\hbar}\int_0^T\int_0^t dt\,dt' \int_0^\infty d\omega\,\omega\sin(\omega(t-t'))\left(X(t)-X'(t)\right)\left(X(t')+X'(t')\right)\Biggr]\ .
 \label{bigexpression}
\ea
Here we have inserted the vacuum influence phase~\eqref{ifvaclong}. Recall that $X(t)= \frac{d^2}{dt^2} \xi^2(t)$ and $X'(t) = \frac{d^2}{dt^2} \xi'^2(t)$. We again observe that, in a coherent state, the action for $\xi$ acquires a piece corresponding to the interaction with a classical gravitational wave $\bar{h}$. The last two terms arise from $F_0$ and encode the vacuum fluctuations of the gravitational field. We now analyze these two terms in further detail; we shall see that they are related to fluctuation and dissipation.
 
 \subsection*{Dissipation}
 
 \noindent
Consider the last term in the exponent in~\eqref{bigexpression}. The integral over $\omega$ can be evaluated by using the distributional identity
\be
\frac{1}{\pi}\int_0^\infty d\omega\,\omega\sin(\omega(t-t'))=-\delta'(t-t')\ ,
\ee
where $\delta'$ is the derivative of the Dirac delta function with respect to its argument. Then
 \ba
 &&\hspace{-1cm}\frac{im_0^2 G}{4\pi\hbar}\int_0^T\int_0^t dt\,dt' \int_0^\infty d\omega\,\omega\sin(\omega(t-t'))\left(X(t)-X'(t)\right)\left(X(t')+X'(t')\right)\nn\\
 &=&-\frac{im_0^2 G}{4\hbar}\int_0^T\int_0^t dt\,dt'\,\delta'(t-t')\left(X(t)-X'(t)\right)\left(X(t')+X'(t')\right)\nn\\
 &=&-\frac{im_0^2 G}{8\hbar}\int_0^Tdt\,\left(X(t)-X'(t)\right)\left(\dot{X}(t)+\dot{X}'(t)\right)\nn\\
 &&+\frac{im_0^2G}{4\hbar}\int_0^T dt\,\delta(0)\left(X(t)^2-X'(t)^2\right)-\frac{im_0^2G}{8\hbar}\left(X(0)^2-X'(0)^2\right)\ .
 \ea
The last term vanishes as a consequence of the boundary conditions in the path integral, as mentioned after~\eqref{transfinal}. The penultimate term, while divergent, takes the form of a difference of actions and can therefore be cancelled through the addition of an appropriate counterterm to the free particle action.  This leaves us with the first term, which contains third-order derivatives of $\xi$ and $\xi'$. This remaining term cannot be expressed as a difference of actions and, consistent with this, we will see shortly that it leads to dissipative dynamics for $\xi$.

\subsection*{Fluctuation}

\noindent
Let us turn now to the second-last term in the exponent in~\eqref{bigexpression}. Using its symmetry to change the limits on the integrals, we can write it as
\be
-\frac{m^2_0}{32\hbar^2}\int_0^T\int_0^T dt\,dt'\,A_0(t,t')\left(X(t)-X'(t)\right)\left(X(t')-X'(t')\right)\ ,
\ee
where we have defined
\be
A_0(t,t')\equiv \frac{4\hbar G}{\pi}\int_0^\infty d\omega\,\omega\cos(\omega(t-t'))\,.
\label{aknot}
\ee
Although $A_0$ is formally divergent, we can imagine that it is regulated in some manner;  for example one could impose a hard cutoff because our formalism surely does not hold for frequencies higher than the Planck scale. Alternatively we can also view $A_0$ as a distribution
\be
A_0(t-t')=-\frac{4\hbar G}{\pi}\mathcal{H}\frac{1}{(t-t')^2}
\ee
where the Hadamard finite-part distribution $\mathcal{H}\frac{1}{x^2}$ is defined when integrated against a test function $\phi(x)$ by
\be
\int_{-\infty}^{\infty}dx\,\phi(x)\,\mathcal{H}\frac{1}{x^2}\equiv \int_{-\infty}^{\infty}dx\,\frac{\phi(x)-\phi(0)-x\phi'(0)}{x^2}\,.
\ee

To proceed, we employ a clever trick due to Feynman and Vernon. We note that the exponential term involving $A_0$ can be expressed as a Gaussian path integral over an auxiliary function ${\cal N}_0(t)$:
\ba
&&\exp\left[-\frac{m^2_0}{32\hbar^2}\int_0^T\int_0^T dt\,dt'\,A_0(t,t')\left(X(t)-X'(t)\right)\left(X(t')-X'(t')\right)\right]\nn\\
&&\hspace{-10mm}
=\int {\cal D}{\cal N}_0\exp\left[-\frac{1}{2}\int_0^T\int_0^Tdt\,dt'\,A_0^{-1}(t,t'){\cal N}_0(t){\cal N}_0(t')+\frac{i}{\hbar}\int_0^T dt\frac{m_0}{4}{\cal N}_0(t)\left(X(t)-X'(t)\right)\right]\,.
\label{funcgauss}
\ea
Here $A_0^{-1}$ is the operator inverse of $A_0$, formally obeying $\int_0^Tds A_0(t,s)A_0^{-1}(s,t')=\int_0^Tds A_0^{-1}(t,s)A_0(s,t')=\delta(t-t')$. Equation~\eqref{funcgauss} has an elegant interpretation. The function ${\cal N}_0(t)$ is evidently a stochastic (random) function with a Gaussian probability density. (An overall normalization factor has been absorbed in the measure.) Moreover the stochastic average of ${\cal N}_0(t)$ clearly vanishes:
\be
\langle {\cal N}_0(t) \rangle \equiv  \int {\cal D}{\cal N}_0\exp\left[-\frac{1}{2}\int_0^T\int_0^Tdt\,dt'\,A_0^{-1}(t,t'){\cal N}_0(t){\cal N}_0(t')\right]{\cal N}_0(t)=0\ . 
\ee
Thus ${\cal N}_0(t)$ is naturally interpreted as noise. We can also then see that $A_0$ is the auto-correlation function of ${\cal N}_0(t)$ since 
\be
\langle {\cal N}_0(t){\cal N}_0(t') \rangle \equiv  \int {\cal D}{\cal N}_0\exp\left[-\frac{1}{2}\int_0^T\int_0^Tdt\,dt'\,A_0^{-1}(t,t'){\cal N}_0(t){\cal N}_0(t')\right]{\cal N}_0(t){\cal N}_0(t')=A_0(t,t')\ .
\ee
The auto-correlation $A_0$ fully describes the properties of the noise ${\cal N}_0(t)$ as, by Wick's theorem, any higher moment is expressible in terms of sums of products of $A_0$. 

The upshot of the Feynman-Vernon trick is that we are able to transform a term that coupled $\xi$ and $\xi'$ into one that can be written as a difference of two actions. Furthermore, the new actions now contain an external function ${\cal N}_0(t)$ which, as we have seen, has the interpretation of noise. We can analyze this noise further by examining the auto-correlation function. First note from~\eqref{aknot} that, because $A_0(t,t')$ depends only on $\tau=t-t'$, the noise must be stationary. Observe also that $A_0$ is symmetric under $\tau \rightarrow -\tau$. Then taking the Fourier transform of the auto-correlation function yields the power spectrum of the noise:
\be
S(\omega)\equiv \int_{-\infty}^\infty d\tau e^{-i\omega\tau}A_0(\tau) = 4G\hbar\omega\ .
\ee
As is manifest from the presence of $\hbar$, this is a fundamental  noise of quantum origin. Moreover, the $\omega$-dependence indicates that it is not white noise, but rather correlated noise with a characteristic spectrum.

\subsection*{Effective dynamics of the arm length}

\noindent Putting all this together, we find that the transition probability~\eqref{bigexpression} can be written as
\ba
P_{\bar{h}}(\phi_A \to \phi_B)
&\equiv& \int d \xi_i d \xi'_i d \xi_f d \xi'_f \, \phi^*_A(\xi'_i)  \phi_B(\xi'_f) \phi^*_B(\xi_f) \phi_A(\xi_i)\times\nn\\
&&\hspace{-15mm}\int \tilde{\cal D}\xi \tilde{\cal D}\xi' {\cal D}{\cal N}_0\exp\left[-\frac{1}{2}\int_0^T\int_0^Tdt\,dt'\,A_0^{-1}(t-t'){\cal N}_0(t){\cal N}_0(t')\right]\times\nn\\
&&\hspace{8mm}\exp\Biggr[\frac{i}{\hbar} \int_{0}^{T} dt \left\{\frac{1}{2} m_0 \left(\dot{\xi}^2 - \dot{\xi}'^{2}\right)+\frac{1}{4}m_0\left(\bar{h}(t)+{\cal N}_0(t)\right)\left(X(t)-X'(t)\right)\right\}\nn\\
&&\hspace{12mm}-\frac{im_0^2 G}{8\hbar}\int_0^Tdt\,\left(X(t)-X'(t)\right)\left(\dot{X}(t)+\dot{X}'(t)\right)
\Biggr]\ .
\label{finalprob}
\ea
We now have a triple path integral as the noise function ${\cal N}_0(t)$ comes with its own Gaussian probability measure; indeed we can view the path integral over ${\cal N}_0(t)$ as a stochastic average of the last exponent. Notice also that the noise ${\cal N}_0(t)$ adds to the classical gravitational wave $\bar{h}(t)$. Finally the term in the last line precludes us from regarding the quantum effects of the vacuum fluctuations as arising from an effective action; as mentioned earlier that term does not separate into the form $\frac{i}{\hbar}(S[\xi]-S[\xi'])$.

We have calculated the exact transition probability for the arm length to go from an initial quantum state $|\phi_A\rangle$ to a final one $|\phi_B\rangle$. But we expect that the arm length $\xi$ -- which can also be regarded as the position of a macroscopic mass $m_0$ -- is essentially a classical degree of freedom. Consequently, the $\xi$ and $\xi'$ path integrals in~\eqref{finalprob} should be dominated by the contribution of their saddle points. These are determined by paths $\xi(t)$, $\xi'(t)$ obeying two coupled differential equations:
\be
\ddot{\xi}-\frac{1}{2}\left[\ddot{\bar{h}}+\ddot{{\cal N}}_0-\frac{m_0G}{2}\left(\dddot{X}+\dddot{X}'-\ddot{X}+\ddot{X}'\right)\right]\xi=0\ ,
\label{eomvarxi}
\ee
as well as its counterpart obtained by interchanging $\xi$ and $\xi'$. Generically there are  solutions of this system of coupled differential equations for which $\xi(t)$ and $\xi'(t)$ are different. We will discuss this interesting phenomenon of asymmetric semi-classical paths, which is not specific to gravitational radiation, in a separate publication.   Here we make the simplifying Ansatz that $\xi(t)=\xi'(t)$. Then $X(t)=X'(t)$ and~\eqref{eomvarxi} reduces to the Langevin-like equation
\be
\ddot{\xi}(t)-\frac{1}{2}\left[\ddot{\bar{h}}(t)+\ddot{{\cal N}}_0(t)-\frac{m_0G}{c^5}\frac{d^5}{dt^5}\xi^2(t)\right]\xi(t)=0\ .
\label{langevineq}
\ee
We have restored factors of $c$ and substituted $X$ for its expression in terms of $\xi$, \eqref{Xintermsofxi}. This is our main result; let us discuss it in some detail. The equation describes the quantum-corrected dynamics of the arm length $\xi$ or, equivalently, of the position of the second free-falling mass relative to the first; it is the quantum geodesic deviation equation~\cite{Parikh:shortpaper,Haba:2020jqs}. It contains, within the brackets, three terms that source the relative acceleration $\ddot{\xi}$. The first of these terms is present also in the classical equation~\eqref{classeom}; as before it determines the tidal acceleration due to a background gravitational wave. The remaining two terms correspond to fluctuation and dissipation respectively. The last, non-linear, fifth derivative term is a gravitational radiation reaction term. It is analogous to the Abraham-Lorentz acceleration in electromagnetism. But whereas in the electromagnetic case the radiation reaction term has three derivatives, here there are five derivatives~\cite{Misner:1974qybis,Mino:1996nk,Quinn:1996am}; this is to be expected from the presence of the extra derivative in the gravitational field interaction. In contrast to the electromagnetic case, the gravitational radiation reaction term is non-linear in $\xi$; this non-linearity can be traced to the non-linear interaction term in~\eqref{Somega}. The pathologies that ensue when the Abraham-Lorentz equation of classical electromagnetism is taken literally have been the subject of much confusion.  It has long been anticipated that quantum effects will somehow remedy the situation. Here we see that such equations are approximations to path integrals, \eqref{finalprob}, that are free of pathologies.  Most importantly, \eqref{langevineq} contains a quantum noise ${\cal N}_0(t)$ as a source; the presence of this term means that the equation is in fact a stochastic differential equation. This is intuitively appealing: it conforms to the expectation that a quantum field will induce random fluctuations in any classical degree of freedom it interacts with. This randomness has the effect of altering the dynamics of the classical degree of freedom so that it is necessarily described by a stochastic -- rather than a deterministic -- equation of motion. Notice that this noise is present even in the absence of an accompanying classical gravitational wave. We will discuss the phenomenology of this equation in Section~\ref{discussion}.

\subsection*{Extension to thermal and squeezed vacua}

\noindent In~\eqref{bigexpression} we specialized to the important case of a coherent gravitational field state, $|\Psi\rangle=|\bar{h}\rangle$. Let us now extend the effective dynamics of $\xi$ to other classes of gravitational field states, namely thermal and squeezed states. This is readily done: we have already done the hard work of computing the influence functional for these cases. It is now simply a matter of inserting the influence functional into~\eqref{transfinalfield}, performing the Feynman-Vernon trick, and taking a saddle point.

Consider first a gravitational field in a thermal state. We have already computed the additional influence phase in this state, \eqref{phithermal}; this phase is in addition to the influence phase of the vacuum which is always present. Comparing the additional thermal influence phase with that of the vacuum, \eqref{ifvaclong}, we see that it contains only a real (fluctuation) part and no imaginary (dissipation) part. We can again rewrite the real part using the Feynman-Vernon trick:
\ba
&&\exp\left[-\frac{m^2_0}{32\hbar^2}\int_0^T\int_0^T dt\,dt'\,A_{\rm th}(t,t')\left(X(t)-X'(t)\right)\left(X(t')-X'(t')\right)\right]=\nn\\
&&\hspace{-6mm}
\int {\cal D}{\cal N}_{\rm th}\exp\left[-\frac{1}{2}\int_0^T\int_0^Tdt\,dt'\,A_{\rm th}^{-1}(t,t'){\cal N}_{\rm th}(t){\cal N}_{\rm th}(t')+\frac{i}{\hbar}\int_0^T dt\frac{m_0}{4}{\cal N}_{\rm th}(t)\left(X(t)-X'(t)\right)\right]\ .
\label{funcgausstherm}
\ea
Here, from~\eqref{phithermal}, we read off 
\be
A_{\rm th}(t,t')=  \frac{8\hbar G}{\pi}\int_0^\infty \frac{\omega d\omega}{e^{\frac{\hbar\omega}{k_BT}}-1}\cos(\omega(t-t'))=\frac{4\hbar G}{\pi}\left(\frac{1}{(t-t')^2}-\frac{\pi^2k_B^2T^2}{\hbar^2\sinh^2\left(\frac{\pi k_BT(t-t')}{\hbar}\right)}\right)\,.
\label{Ath}
\ee
Unlike the vacuum auto-correlation function, the thermal auto-correlation function is finite. 
We see from~\eqref{funcgausstherm} that in the thermal state the arm length is subject to an additional Gaussian noise source. The power spectrum of this noise is given by
\be
S_{\rm th}(\omega)=\frac{8\hbar G\omega}{e^{\frac{\hbar\omega}{k_BT}}-1}\ .
\label{spectherm}
\ee
After performing the saddle point over the $\xi$, $\xi'$ path integrals, setting $\xi=\xi'$, and remembering to include the vacuum contributions, we finally have
\be
\ddot{\xi}(t)-\frac{1}{2}\left[\ddot{{\cal N}}_0(t)+\ddot{{\cal N}}_{\rm th}(t)-\frac{m_0G}{c^5}\frac{d^5}{dt^5}\xi^2(t)\right]\xi(t)=0\ .
\label{langevineqtherm}
\ee
This is the Langevin equation for the arm length in the presence of a thermal gravitational field. It contains an additional correlated noise term with power spectrum~\eqref{spectherm}.

Next consider a gravitational field in a squeezed vacuum. The additional influence phase in this state was computed in~\eqref{phisqueezed}. We again see that there is only a real (fluctuation) part which will contribute to the noise. Performing the Feynman-Vernon trick, we find
\be
A_{z}(t,t')=\frac{4\hbar G}{\pi}(\cosh 2r-1)\int_0^\infty d\omega\,\omega \cos(\omega(t-t'))-\frac{4\hbar G}{\pi}\sinh 2r\int_0^\infty d\omega\,\omega\cos(\omega(t+t'))\ .
\label{astat}
\ee
Unlike our previous examples, the noise in the squeezed state is not stationary because $A_z(t,t')$ does not depend only on $t-t'$; indeed, the time-modulation of the noise in squeezed states is a familiar phenomenon in quantum optics~\cite{Walls:1983}. We can decompose $A_z(t,t') =A_{\rm stat}(t-t')+A_{\rm non-stat}(t+t')$ and perform the Feynman-Vernon trick for these two parts separately. This introduces corresponding stationary and non-stationary noises, ${\cal N}_{\rm stat}$ and ${\cal N}_{\rm non-stat}$, and, mutatis mutandi, we find
\be
\ddot{\xi}(t)-\frac{1}{2}\left[\ddot{{\cal N}}_{0}(t)+\ddot{{\cal N}}_{\rm stat}(t)+\ddot{{\cal N}}_{\rm non-stat}(t)-\frac{m_0G}{c^5}\frac{d^5}{dt^5}\xi^2(t)\right]\xi(t)=0\ .
\label{langevineqsqueezed}
\ee
Notice from~\eqref{astat} that, for the idealized uniform squeezing that we have been considering, $A_{\rm stat}$ is proportional to the auto-correlation of the vacuum $A_0(t,t')$, which we had previously calculated in \eqref{aknot}. With suitable redefinitions, we can therefore combine ${\cal N}_0$ and ${\cal N}_{\rm stat}$ into a single stationary noise term $\sqrt{\cosh 2r}{\cal N}_0$. Remarkably, the amplitude of the vacuum noise is enhanced by a factor of $\sqrt{\cosh 2r}$:
\be
\ddot{\xi}(t)-\frac{1}{2}\left[\sqrt{\cosh 2r}\ddot{{\cal N}}_{0}(t)+\ddot{{\cal N}}_{\rm non-stat}(t)-\frac{m_0G}{c^5}\frac{d^5}{dt^5}\xi^2(t)\right]\xi(t)=0\ .
\label{langevineqsqueezedpart}
\ee
This means that if $r\gg 1$, the squeezed vacuum fluctuations lead to an exponential enhancement of the quantum noise in the equation of motion of the arm length; the same result has also been obtained without using influence functionals~\cite{Kanno:2020usf}. The possible effect of squeezed gravitational states on the propagation of photons within LIGO has been discussed recently~\cite{Guerreiro:2019vbq}.

\section{Phenomenology}
\label{discussion}

\noindent Our main result is that the classical geodesic deviation equation is replaced by the Langevin equation~\eqref{langevineq} which is a non-linear stochastic differential equation. We therefore predict the existence of a fundamental noise originating in the quantization of the gravitational field. In order for this noise to be detectable at gravitational wave interferometers, two requirements must be met. First, the amplitude of the noise should not be too small. Second, the noise must be distinguishable from the many other sources of noise at the detector.

Let us begin by estimating the noise amplitude. We will need to make some approximations. The first step is to discard the fifth-derivative radiation reaction term in the Langevin equation. We do this mainly for simplicity, but it seems plausible that if the arm length $\xi$ is measured in some manner that is coarse-grained in time, then its higher derivatives could be negligible. With this approximation, the equation of motion becomes a stochastic Hill equation:
\be
\ddot{\xi}(t)-\frac{1}{2}\left[\ddot{\bar{h}}(t)+\ddot{{\cal N}}(t)\right]\xi(t)=0\ .
\label{hilleq}
\ee
Here ${\cal N}$ stands for any of the noise terms we have considered, and we have also allowed for the possible presence of a classical gravitational background, $\bar{h}$. Next, the linearity of this equation allows us to write the approximate solution as
\be
\xi(t)\approx \xi_0 \left(1+\frac{1}{2}\left(\bar{h}(t)+\mathcal{N}(t)\right)\right)\ ,
\ee
because the resting arm length $\xi_0$ is many orders of magnitude larger than its fluctuations. This equation shows that the fundamental noise ${\cal N}$ induces random fluctuations in the arm length $\xi$. The technology we have developed allows us to calculate the statistical properties of these jitters, such as their mean, standard deviation, auto-correlation function, power spectrum, etc., with the help of the auto-correlation function of ${\cal N}$, viz. $A(t,t')$. 
Since ${\cal N}$ averages to 0, we see from~\eqref{hilleq} that the average value of $\xi$ is, as expected, its classical value:
\be
\langle \xi(t)\rangle \approx \xi_0\left(1+\frac{1}{2}\bar{h}(t)\right)\ .
\ee
Then the standard deviation is
\be
\sigma(t) \equiv \left\langle\left( \xi(t) -\langle\xi(t)\rangle\right)^2 \right\rangle^{\frac{1}{2}}\approx\frac{\xi_0}{2}\sqrt{\langle{\cal N}(t){\cal N}(t)\rangle} = \frac{\xi_0}{2}\sqrt{A(t,t)}\ .
\ee
Let us make some estimates. 

In the case of vacuum fluctuations, $A=A_0$, this quantity is formally divergent; see~\eqref{aknot}. However, the detector is not sensitive to arbitrarily high frequencies. We can crudely approximate the $\omega$ integral appearing in $A_0$ by introducing a cut-off at the highest frequency $\omega_{\rm max}$ to which the detector could be sensitive. Now, our derivation (see comments before~\eqref{refdipole}) relied on a dipole-like approximation; hence $\omega_{\rm max}$ can be estimated by $2\pi c/ \xi_0$, although in practice $\omega_{\rm max}$ is typically lower. Then
\be
\sigma_0 \approx\frac{\xi_0}{2}\sqrt{A_0(t,t)} = \xi_0\, \omega_{\rm max} \sqrt{\frac{\hbar G}{2\pi c^5}} \sim 10^{-35}\text{m}\,.
\ee
This is roughly the scale of the Planck length and about 17 orders of magnitude beyond the technological limits of an experiment such as LIGO. Evidently, detecting vacuum fluctuations in the gravitational field with a gravitational interferometer appears impossible. Nor does including a background gravitational wave help: a more careful estimate (assuming that the stochastic noise can be approximated as an It\^o process) shows that in the presence of a gravitational wave, the quantum noise is enhanced only by a tiny factor of $1+\bar{h}$. This contradicts claims in the literature~\cite{Lieu:2017lzh} according to which graviton shot noise should already have been detected at LIGO.

Next let us consider fluctuations in a thermal state. Then $A=A_{\rm th}$, \eqref{Ath}, and we find a finite expression for the standard deviation of the arm length:
\be
\sigma_{\rm th}(t)\approx \frac{\xi_0}{2}\sqrt{A_{\rm th}(t,t)}=\xi_0\sqrt{\frac{\pi(k_BT)^2G}{3\hbar c^5}}\,.
\label{thermalnoiseover}
\ee
This is a theoretical limit; in practice limits on the detector sensitivity again require that the integral over $\omega$ appearing in $A_{\rm th}$ should be cut off at the highest frequency to which the detector is sensitive (which is typically well below the frequency of the peak of the Planck distribution, $\hbar\omega_{\rm max}\ll k_BT$). The relevant expression should instead read
\be
\sigma_{\rm th}\approx\xi_0\sqrt{\frac{2G k_BT\omega_{\rm max}}{\pi c^5}}\ .
\label{thermalnoise}
\ee
For LIGO ($\xi_0\sim 1\text{km}$, $\omega_{\rm max}\sim 10^6\,\text{rad\,s}^{-1}$), the noise due to the isotropic cosmic gravitational wave background ($T\sim 1\,\text{K}$)   yields a $\sigma_{\rm th}$ of order $10^{-31}\text{m}$ or about 13 orders of magnitude beyond its current technological limits. For LISA ($\xi_0\sim 10^6\,\text{km}$, $\omega_{\rm max}\sim 1\,\text{rad\,s}^{-1}$), the situation would be slightly improved with a noise level of order $10^{-28}\text{m}$, ``only'' 10 orders of magnitude beyond its projected sensitivity. Notice that using~\eqref{thermalnoiseover} instead of~\eqref{thermalnoise} would overestimate the noise amplitude by about 3 orders of magnitude for LIGO and 5 for LISA; most of the power in the thermal noise is concentrated at high frequencies that are inaccessible to LIGO (and even more so to LISA). We can also consider gravitational fields due to localized thermal sources, such as evaporating black holes. Here in principle, the temperature can be much higher, as could be expected for exploding primordial black holes. However, as discussed earlier, the quantum noise contribution would be suppressed by a tiny geometric factor of $\frac{1}{4}\left(\frac{r_S}{r}\right)^2$ where $r_S$ is the black hole's Schwarzschild radius and $r \gg r_S$ its distance from the detector. It might be worthwhile to check whether there are regions of the parameter space of primordial black hole density distributions for which the collective background of evaporating black holes might allow for a detectable signal.

Perhaps the most intriguing prospect is the quantum noise from a squeezed vacuum. In this case, as discussed in the previous section, the noise has both stationary and time-dependent components. Focusing on the stationary piece, we find that
\be
\sigma_{\rm squeezed}= \sigma_0 \sqrt{\cosh 2r} \ .
\ee
For large values of $r$, the squeezing results in exponential enhancement of the fluctuation of the detector arm length, as also found in \cite{Kanno:2020usf}. It would be very interesting to see whether there are realistic physical sources of the gravitational field that could yield squeezed states with values of $r$ for which the noise might be detectable.

Finally, it is worth emphasizing that the fundamental noise arising from the quantization of the gravitational field has some particular properties that could potentially help to distinguish it from other, more mundane, sources of noise. Indeed, it is non-transient (even stationary in some cases) and, for many classes of quantum states, its precise power spectrum is analytically calculable. Furthermore, the noise is likely to be correlated between nearby detectors. To see this, consider an additional detector degree of freedom $\zeta(t)$. Schematically, this adds a term of the form $-\frac{1}{2}m_0\dot{\bar{h}}\dot{\zeta}\zeta$ to the interaction Lagrangian which effectively replaces $X$ by $X+Y$ in the influence functional~\eqref{geninffunc}, where $Y(t)=\frac{d^2}{dt^2}\zeta^2(t)$. Then the identity~\eqref{funcgauss} results in a {\em single} stochastic function ${\cal N}$ multiplying both $X$ and $Y$, leading to correlated noise between the two detectors. 

\section{Summary}
\label{summary}

\noindent In this paper we have considered Einstein gravity coupled to a model gravitational wave detector. We quantized the resultant theory and integrated out the gravitational field to obtain the Feynman-Vernon influence functional. This encompassed the effect on the detector of its coupling to a quantized gravitational field; we calculated the influence functional for a variety of quantum states of the gravitational field. We then arrived at a stochastic equation of motion for the length of the detector arm which generically contains a quantum noise term. We identified the power spectrum of the noise and estimated its amplitude for various states. For vacuum, or for sources for which it is valid to employ deterministic equations and treat the gravitational field linearly, we found only minuscule corrections to the classical treatment. But other sources can produce gravitational field states, including quasi-thermal and squeezed states, that are more promising in this regard.

Unlike in electrodynamics, where the linearity of the theory prevents sources governed by deterministic dynamics from producing non-coherent states~\cite{Glauber:1963tx}, the non-linearity of general relativity naturally gives rise to non-coherent states when strong-field effects become significant, as during the end stages of binary black hole mergers. The calculation of these states and their properties is worthy of further attention. Finally, we also identified interesting issues relating to the generally asymmetric nature of semi-classical paths in influence functionals and the formal origin of radiation reaction. These topics are under active investigation.

\bigskip
\noindent
{\bf Acknowledgments}\\
\noindent
We thank Paul Davies, Bei-Lok Hu, Phil Mauskopf, and Tanmay Vachaspati for conversations. During the course of this work, MP and GZ were supported in part by John Templeton Foundation grant 60253. GZ also acknowledges support by Moogsoft. FW is supported in part by the U.S. Department of Energy under grant DE-SC0012567, by the European Research Council under grant 742104, and by the Swedish Research Council under contract 335-2014-7424.

\end{document}